\documentclass{aa}
\pdfoutput=1
\usepackage{amsmath}
\usepackage{graphicx}
\usepackage[caption=false]{subfig}
\usepackage{appendix}
\usepackage{threeparttable}
\usepackage{txfonts}
\usepackage{color}
\usepackage{float}
\usepackage[version=4]{mhchem} 

\usepackage{natbib}
\bibpunct{(}{)}{;}{a}{}{,} 

\usepackage{epstopdf}

\usepackage[off]{auto-pst-pdf}
\begin{document}

\title{Spectroscopic measurements of \ce{CH3OH} in layered and mixed interstellar ice analogues\thanks{Based on experiments conducted at the Max-Planck-Institut f\"ur Extraterrestrische Physik, Garching.}\thanks{The raw spectra are available in electronic form at the CDS via anonymous ftp to cdsarc.u-strasbg.fr (<...>) or via http://cdsweb.u-strasbg.fr/<...>}}

\author{B. M\"uller \inst{\ref{MPE}}
        \and
        B. M. Giuliano \inst{\ref{MPE}}
        \and
        M. Goto \inst{\ref{USM}}
        \and
        P. Caselli \inst{\ref{MPE}}}

\offprints{Birgitta M\"uller}

\institute{Max-Planck-Institut f\"ur extraterrestrische Physik, Gie{\ss}enbachstrasse 1, 85748 Garching bei M\"unchen, Germany \label{MPE}\\
       \email{bmueller@mpe.mpg.de}
           \and
           Universit\"ats-Sternwarte M\"unchen, Scheinerstr. 1, D-81679 M\"unchen, Germany \label{USM}\\}

\date{Received 8 October 2020 / Accepted 21 June 2021}
\authorrunning{B. M\"uller et al.}  
\titlerunning{Layers}

\abstract
{The molecular composition of interstellar ice mantles is defined by gas-grain processes in molecular clouds, with the main components being \ce{H2O}, CO, and \ce{CO2}. Methanol (\ce{CH3OH}) ice is detected towards the denser pre-stellar cores and star-forming regions, where large amounts of CO molecules freeze out and get hydrogenated on top of the icy grains. The thermal heating from nearby protostars can further change the ice structure and composition. Despite the several observations of icy features carried out towards molecular clouds and along the line of site of protostars, it is not yet clear if interstellar ices are mixed or if they have a layered structure.}
{We aim to examine the effect of mixed and layered ice growth in dust grain mantle analogues, with specific focus on the position and shape of methanol infrared bands, so dedicated future observations could shed light on the structure of interstellar ices in different environments.}
{Mixed and layered ice samples were deposited on a cold substrate kept at a temperature of 10~K using a closed-cycle cryostat placed in a vacuum chamber.
The spectroscopic features were analysed by Fourier transform infrared spectroscopy. Different proportions of the most abundant four molecular species in ice mantles, namely \ce{H2O}, \ce{CO}, \ce{CO2,} and \ce{CH3OH}, were investigated, with a special attention placed on the analysis of the \ce{CH3OH} bands.}  
{We measure changes in the position and shape of the \ce{CH} and \ce{CO} stretching bands of \ce{CH3OH} depending on the mixed or layered nature of the ice sample.
Spectroscopic features of methanol are also found to change due to heating.}
{A layered ice structure best reproduces the \ce{CH3OH} band position recently observed towards a pre-stellar core and in star-forming regions. Based on our experimental results, we conclude that observations of \ce{CH3OH} ice features in space can provide information about the structure of interstellar ices, and we expect the James Webb Space Telescope (JWST) to put stringent constraints on the layered or mixed nature of ices in different interstellar environments, from molecular clouds to pre-stellar cores to protostars and protoplanetary discs.}


\maketitle

\section{Introduction}
\label{intro}
\indent\indent The interpretation of astronomical observations relies on the contribution provided by laboratory data and theoretical modelling. 
The spectroscopic features of interstellar ice analogues recorded in the laboratory at infrared wavelengths provide important information to shed light on the composition and physical state of the icy mantles of interstellar dust grains.
Many decades of laboratory investigations dedicated to the study of astrophysical relevant ice samples have offered a rich bibliography that makes information available for the most commonly observed ice mixtures, which can be found in commonly known databases.

\citet{2011ApJ...740..109O} showed that the most realistic picture of ice mantles covering the dust grains is a layered model, in which the chemical composition of the ice changes as the ice accretes on the dust surface following the changes in the gas-phase chemical composition.
This produces changes of the observed ice features depending on the environment probed.
In \citet{Bottinelli2010}, a comparison of low-mass young stellar objects observations (using the \textit{Spitzer} InfraRed Spectrograph) with laboratory studies of ices containing \ce{NH3} and \ce{CH3OH} indicates that \ce{CH3OH} ice is present mostly in pure form or mixed with CO and \ce{CO2}.
The bombardment of icy mantles by energetic particles such as cosmic rays could already affect the ice structure \citep[e.g.][]{Leger_1985,Dartois2020,Ivlev_2015} in the pre-stellar phase, while protostellar activity (thermal heating and shocks) may also provide other mechanisms to mix a previously layered icy mantle onto dust grains.

In the past years, experiments on layered ice analogues have focussed on a number of components that are present in interstellar ices such as \ce{H2O}, CO, \ce{CO2}, \ce{CH4}, HCOOH, or \ce{CH3OH}.
First laboratory studies by \citet{doi:10.1021/jp9710291}, who compared pure water ice, water layered on top of Ar and Ne, as well as \ce{H2O}-Ar, \ce{H2O}-Ne and \ce{H2O}-CO ice mixtures, showed that the spectroscopic parameters are sensitive to structural modifications in the solid ice layers as well as sudden changes in the temperature.

Investigations on layered and mixed CO containing ice analogues by \citet{10.1111/j.1365-2966.2004.08038.x} and \citet{Fraser2004} support the assumption that interstellar CO ice is present in mostly pure layers instead of mixed ices.
Measurements of CO layered on top of other species conducted by Collings and McCoustra \citep{Collings2003} observed entrapment of CO in pores of \ce{H2O} ice as well as a collapse of those pores during heating. Subsequent sub-monolayer coverage experiments characterised the band shifts of CO on top of crystalline and amorphous solid water \citep{Collings2005} and on top of \ce{^{13}CO}, \ce{CO2}, \ce{NH3}, \ce{CH3OH,} and \ce{H2O} \citep{Collings2014}.
The results of \citet{Collings2014} for CO-\ce{CO2} ices agreed well with those of \citet{Broekhuizen_2006}, who measured changes and shifts in the CO and \ce{CO2} bands of pure, mixed, and layered ice analogues. 
\citet{C3CP53767F} examined the structure and dynamics of \ce{H2O} and \ce{CO2} in pure, mixed, and layered ices and they were able to follow \ce{CO2} segregation, which affected the shape and position of the \ce{H2O} bands.
While experiments of \ce{CH3OH} and \ce{C2H5OH} deposited on \ce{H2O} by  Wolff and Brown showed mixing and subsequent co-desorption during heating \citep{Burke2008, Wolff2007}, \citet{Bahr_2008} observed that layered and reverse-layered depositions of \ce{CH3OH} and \ce{D2O} can affect the morphology of the top layers during annealing.
Similarly to our experiments presented in the following sections, \citet{Dawes2016} measured band shifts of mixed and pure \ce{CH3OH} and \ce{H2O} ices, and their results agree well with our findings.

Many of the experimental works mentioned above focus on layered deposition and a comparison of layered and mixed ices is not always provided.
Also, in none of the described experiments have more than two components been analysed, and some examine the ices at temperatures that are too high for comparison with, for example, dense molecular clouds of temperatures between 10 and 20 K.
Only recently, \citet{Ciaravella_2020} experimentally examined X-ray processing of two layered ices where \ce{CH3OH} and CO were deposited on top of an \ce{H2O}:\ce{CH4}:\ce{NH3} ice at 12 K. They observed conversion of hydrogenated species and formation of new molecules both in the bottom and top layers.

More elaborated examination of the influence of pure and mixed ice layers on the spectral signatures is needed.
Here, we focus our attention on solid \ce{CH3OH} features, as methanol has been detected in the gas phase of starless and pre-stellar cores
\citep[e.g.][]{Tafalla_2004,Tafalla_2006,Bizzocchi_2014,Chacon-Tanarro_2019,Scibelli2020} and its distribution was recently modelled by \citet{2017ApJ...842...33V}. The gas-grain chemical code of \citet{2017ApJ...842...33V} predicts that gas-phase \ce{CH3OH} is produced by the reactive desorption of surface methanol, produced on CO-rich ices at the location of the catastrophic CO freeze-out \citep[e.g.][]{Caselli_1999}; a significant fraction of solid \ce{CH3OH} is also predicted to accumulate on the ice mantle. In fact, \citet{Perotti2020} and \citet{Goto_2020} recently detected solid methanol in the direction of the Serpens low-mass star-forming region and the pre-stellar core L1544, respectively.

While \citet{Perotti2020} had difficulties fitting the red wing of the L band between 3.0-3.7 $\mu$m because of a low signal-to-noise (S/N) ratio, \citet{Goto_2020} found an interesting shift in the frequency of the 3.54 $\mu$m \ce{CH3OH} stretching band detected towards the pre-stellar core L1544 when compared to experimental data of ice mixtures examined by \citet{1993ApJS...86..713H}.
With an offset of + 0.01 $\mu$m, this observed methanol band differs from the experimentally measured peak wavelength of 3.53 $\mu$m.
A similar behaviour was found by \citet{Dartois_1999}; while their observations of the protostars RAFGL7009S and W33A mark the methanol CH stretching band at 2827 cm$^{-1}$, laboratory data from \citet{Ehrenfreund_1998} presenting ice mixtures containing methanol show optical depth maxima at positions > 2830 cm$^{-1}$.
Moreover, \citet{Penteado_2015} present observations of methanol ice towards the young stellar object AFGL 7009S, showing a peak position at 3.53 $\mu$m instead of the actual absorption band maximum expected at 3.54 $\mu$m.
They offer an explanation for this discrepancy stating that either a gradient of different ice compositions in the mantles exists or that mantles change in composition along the line of sight.
As presented in the following sections, our experiments suggest a different explanation to the shifts in the $\nu_1$ methanol band position observed by \citet{Dartois_1999}, \citet{Goto_2020}, and \citet{Penteado_2015}: a layered ice structure.

We conducted experiments on ices with compositions resembling that of interstellar ices, where the most abundant species are \ce{H2O}, \ce{CO,} and \ce{CO2}, and \ce{CH3OH} is a minor component, following observations of \citet{2011ApJ...740..109O} in different environments. We provide accurate spectroscopic measurements of ice features in layered and mixed ices containing \ce{CH3OH}, with the aim of assisting the interpretation of the above-mentioned observations of solid methanol.
Moreover, we present the effect of heating on the shape of the vibrational bands in layered and mixed ices. We show that the layered or mixed structure can be deduced from the characteristic features of the observed bands.

The experimental methods employed are described in Sect. 2, followed by the presentation of the results in Sect. 3. A discussion about the measurements can be found in Sect. 4, while Sect. 5 addresses our conclusions.

\section{Experimental methods}\label{expe}
\indent\indent The experiments were performed using CASICE, the cryogenic set-up developed at the Center for Astrochemical Studies (CAS) located at the Max Planck Institute for Extraterrestrial Physics in Garching (Germany). The set-up is composed of a closed-cycle He cryostat purchased from Advanced Research Systems, coupled with a Bruker Fourier Transform Infrared (FTIR) spectrometer.
A detailed description of the set-up can be found in \citet{2018A&A...620A..46M}.

\subsection{Cryogenic set-up}
\label{cryo}
\indent\indent The cryostat is mounted in a stainless steel vacuum chamber of adequate dimensions to be hosted in the sample compartment of the spectrometer. A final vacuum of $10^{-7}$ mbar and a temperature of 10 K is attained after cooling.

The IR spectra were recorded in the 4800-500 cm$^{-1}$ (2.1-20 $\mu$m) frequency range using a standard deuterated triglycine sulfate (DTGS) detector. A spectral resolution of 1 cm$^{-1}$ has been used and the signal has been averaged over 128 scans, as is common practice. No significant differences were noted with spectra taken at higher resolutions \citep[see][]{Broekhuizen_2006, Gerakines2015}.
To obtain a good transmittance in the spectral range, KBr is used in these experiments as optical material for the substrate, while the vacuum chamber optical windows are made of ZnSe.

\subsection{Ice preparation}
\label{ice}
\indent\indent The ice layers are prepared by expansion of suitable gaseous samples into the vacuum chamber followed by condensation on top of the cold substrate. A two-fold gas inlet was used to allow the gases into the CASICE vacuum chamber. The pure and mixed gases used for the ice layers' formation were connected to the two inlets and expanded in the vacuum chamber separately to allow the formation of a layered structure. No contamination from residual gas in the inlet pipe was observed.

The total thickness of the ice samples was kept below 1 $\mu$m to be as close as possible to the typical thicknesses for icy mantles in dense clouds and pre-stellar cores \citep[e.g.][]{2017ApJ...842...33V}. In our set-up, we reached this thickness after several minutes of gas deposition. The contamination from residual water vapour contributes to an ice growth of four monolayers per hour and can be neglected.

Gas mixtures were prepared in glass bulbs using a separate gas line equipped with flow controllers and precision valves. Following the ideal gas law, the molecular ratio in the gas phase was controlled by measuring the partial pressure of each species.

After deposition, the relative abundance of the molecular species in the ice samples were calculated by converting the integrated absorption features into column density values by using band strength values tabulated in \citet{1993ApJS...86..713H} and \citet{1995A&A...296..810G}.

\section{Results}
\label{results}
\indent\indent In this work, we compare pure ices with layered and mixed ices and show the differences in the spectral features. The purpose is to investigate the change in band shape and position of the most prominent methanol features in mixed and layered ices.

In the following three sections, the results of three different sets of experiments are shown. The results of the heating and inverse deposition of selected experiments is presented in Sects. 3.4 and 3.5.
The complete list of performed experiments is shown in Table~\ref{exps}, together with the estimated molecular composition and thickness expressed as number of monolayers.
The chosen abundances of \ce{CH3OH}, \ce{CO,} and \ce{CO2} relative to \ce{H2O} are based on median values obtained in observations of cloud cores and high-mass protostars conducted by \citet{2011ApJ...740..109O}.
All the spectra are presented in raw format, without baseline correction, in order to avoid the appearance of spurious features due to the correction procedure.

\begin{table*}[!htp]
\centering
\begin{threeparttable}
\caption{List of experiments performed in the present work. The number of monolayers has an error of 20\% - 30\%.}
\label{exps}
\begin{tabular}{c c c c c c c c}
\hline \hline \\[-1ex]
Experiment &       & $\chi_{\ce{H2O}}$\tnote{a} & $\chi_{\ce{CH3OH}}$\tnote{a} & & $\chi_{\ce{CO}}$\tnote{a} & $\chi_{\ce{CO2}}$\tnote{a} & Thickness (MLs) \\[1ex]
\hline \\
Pure \ce{H2O}    & & 100\% & --    &               & --   & --   & 1739 \\
Pure \ce{CH3OH}  & & --    & 6\%\tnote{b} &        & --   & --   & 103 \\
1 & 1$^{st}$ layer & 100\% & --    &               & --   & --   & 2344 \\
  & 2$^{nd}$ layer & --    & 5\%   &               & --   & --   & 116 \\
2 & mixed          & 100\% & 4\%   &               & --   & --   & 1530 \\
3 & 1$^{st}$ layer & 100\% & --    &               & --   & --   & 1598 \\ 
  & 2$^{nd}$ layer & --    & 3.5\% & 7\%\tnote{c}  & 50\% & --   & 850 \\ 
4 & 1$^{st}$ layer & 100\% & --    &               & --   & --   & 1352 \\ 
  & 2$^{nd}$ layer & --    & 5\%   & 16\%\tnote{c} & 30\% & --   & 470 \\ 
5 & 1$^{st}$ layer & 100\% & --    &               & --   & --   & 1453 \\ 
  & 2$^{nd}$ layer & --    & 10\%  & 32\%\tnote{c} & 31\% & --   & 596 \\ 
6 & 1$^{st}$ layer & 100\% & --    &               & --   & 39\% & 489 \\ 
  & 2$^{nd}$ layer & --    & 5\%   & 16\%\tnote{c} & 32\% & --   & 351 \\
7 & mixed          & 100\% & 4\%   & 14\%\tnote{c} & 28\% & 34\% & 1650 \\
8 & 1$^{st}$ layer & 100\% & --    &               & --   & --   & 1598 \\
  & 2$^{nd}$ layer & --    & 10\%  & 9\%\tnote{c}  & 113\%& --   & 1962 \\
9 & 1$^{st}$ layer & --    & 5\%   &               & --   & --   & 134 \\
  & 2$^{nd}$ layer & 100\% & --    &               & --   & --   & 2437 \\[0.5ex]
\hline
\end{tabular}
\begin{tablenotes}
\item [a] Abundance of the molecular species in the ice with respect to \ce{H2O}.
\item [b] Abundance of the molecular species in the ice with respect to the number of monolayers of pure \ce{H2O}.
\item [c] Abundance of methanol in the ice with respect to \ce{CO}.
\end{tablenotes}
\end{threeparttable}
\end{table*}

\subsection{Water and methanol}
\label{wm}
\indent\indent As starting test, ices of pure water and pure methanol were deposited in separate experiments.
The resulting spectra were then summed mathematically to obtain the resulting overlapped spectrum.

\begin{figure*}[!htp]
    \centering
    \vspace{0.5cm}
    \subfloat{\includegraphics[scale=0.9]{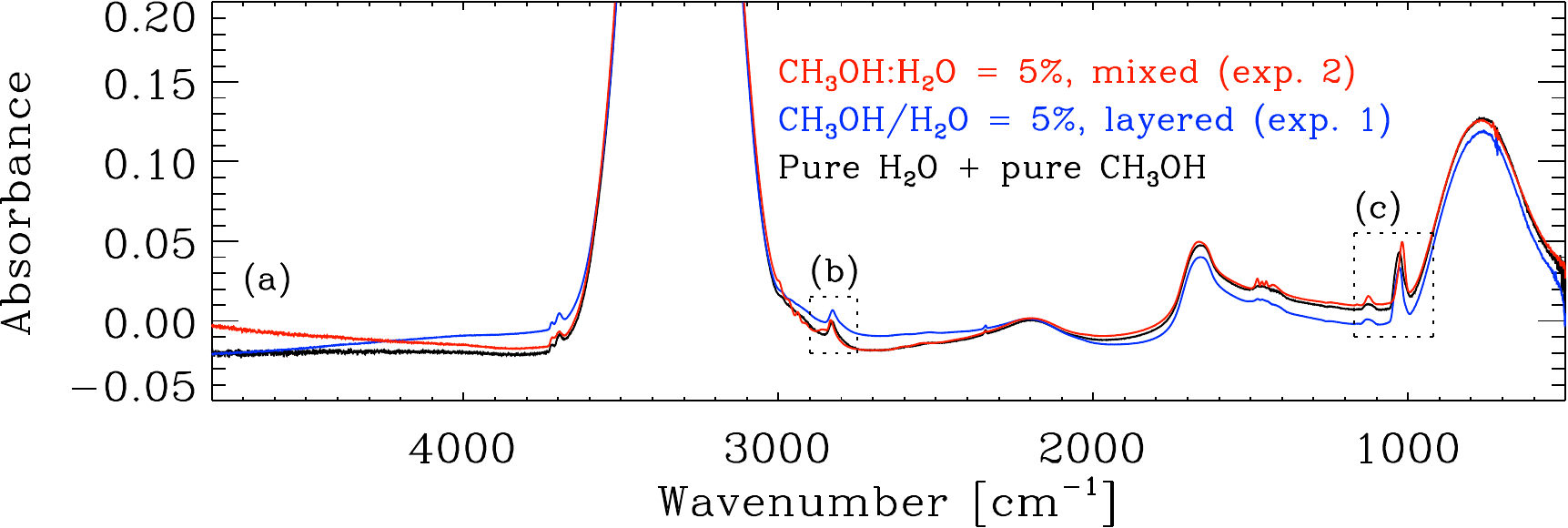}\label{CH3OH_all}}
    
    \subfloat{\includegraphics[scale=.45]{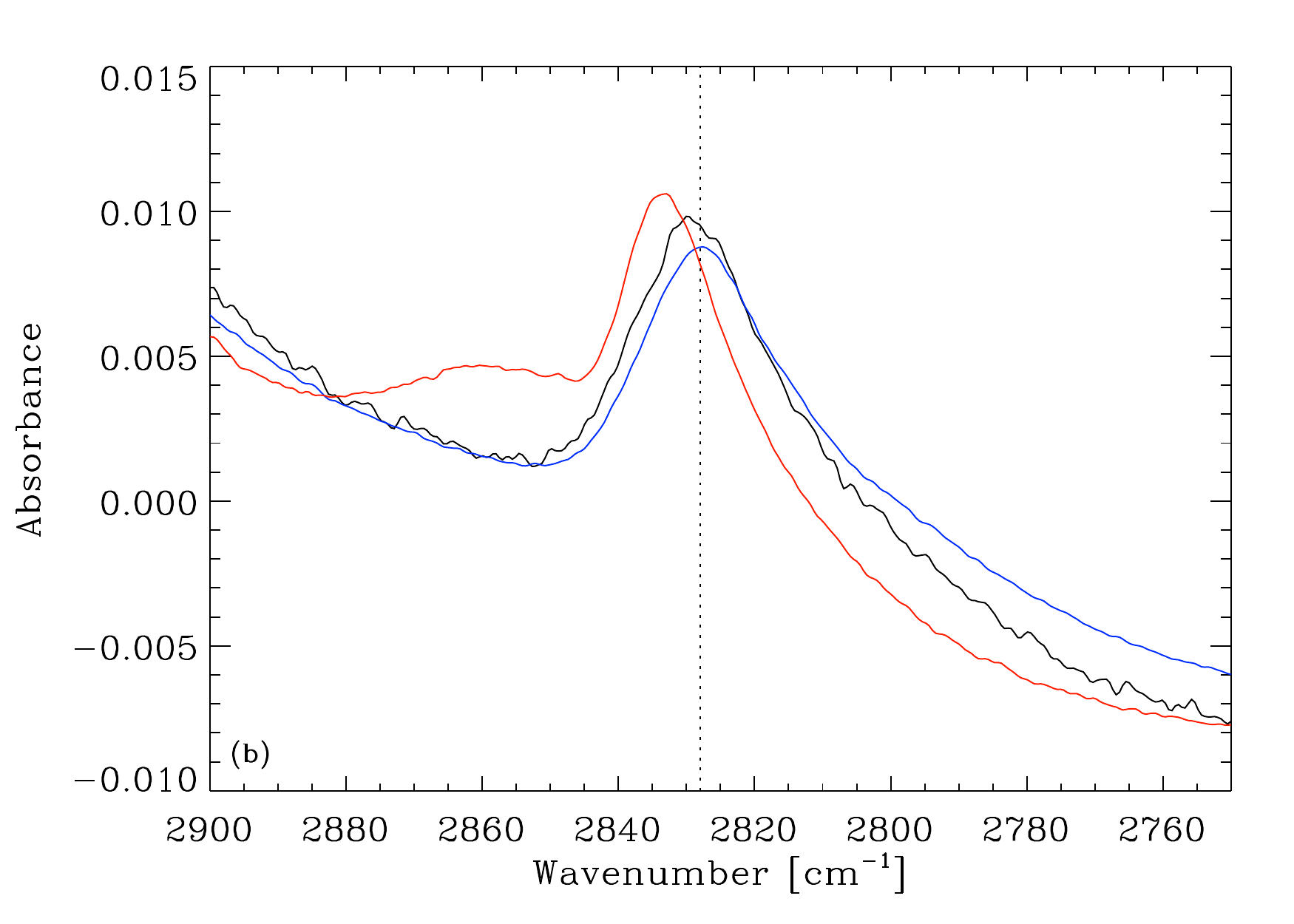}\label{CH3OH_2828}}
    \subfloat{\includegraphics[scale=.45]{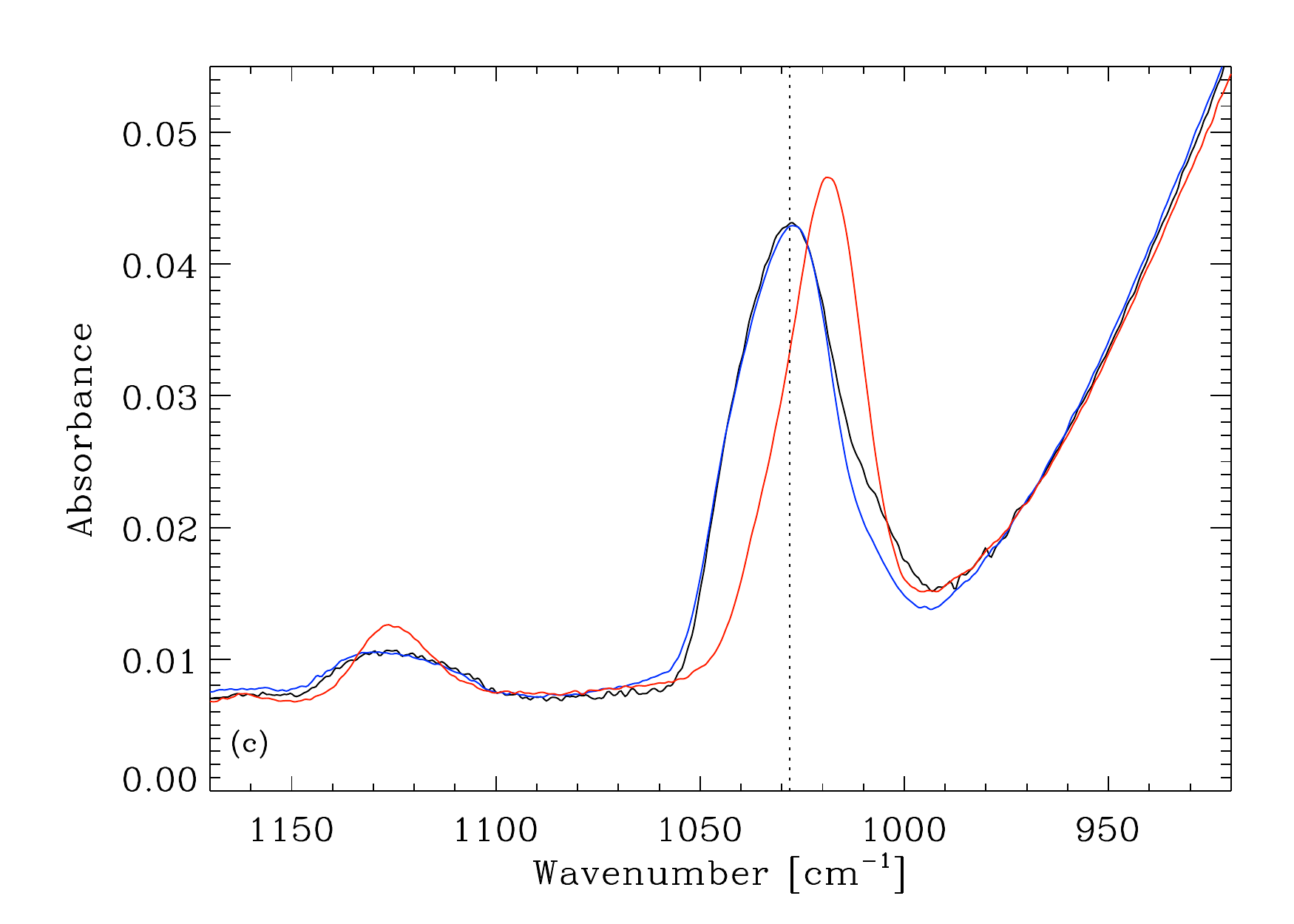}\label{CH3OH_1032}}
    \caption{Spectra of pure methanol and water ices, \ce{CH3OH} on top of \ce{H2O} ice layers (\ce{CH3OH}/\ce{H2O} exp. 1 in Table~\ref{exps}) and \ce{H2O}-\ce{CH3OH} mixture (\ce{CH3OH}:\ce{H2O} exp. 2 in Table~\ref{exps}) in the full frequency range (a), zoomed-in around the 3.54 $\mu$m CH-stretch band (b) and zoomed-in around the 9.75 $\mu$m CO-stretch band (c). Re-scaling was applied to the spectrum of exp. 2, and in Figs.~\ref{CH3OH_2828} and \ref{CH3OH_1032} the spectra are shifted along the y-axis.}
    \label{CH3OH}
\end{figure*}

Similarly to \citet{Dawes2016}, who compared pure and mixed ices, the overlapped spectrum was then compared to the spectra from the two different experiments in which the ice is deposited in layers or pre-mixed in the gas phase (exps. 1 and 2 in Table~\ref{exps}). The results are shown in Fig.~\ref{CH3OH}, where sub-figures (b) and (c) present not only changes in the shape of the band profile but also clearly distinguishable shifts of the CH-stretch band as well as the CO-stretch band maxima position for mixed and layered ices in more detail.

The molecular ratio in the pre-mixed gas sample was chosen in order to match the relative abundance of water and methanol in the layered ice assuming that the ratios are similar in gas and solid phases.
In Fig. \ref{CH3OH}, re-scaling was applied to the spectrum of the mixed \ce{CH3OH}-\ce{H2O} ice.
Moreover, in Figs.~\ref{CH3OH_2828} and \ref{CH3OH_1032} the spectra were shifted along the y-axis for better comparability of the bands.
The re-scaling does not alter the shape of the spectrum and was chosen to guide the eye in the comparison.
Table~\ref{bands} presents the $\nu_3$ CH-stretch and $\nu_8$ CO-stretch absorption band positions of \ce{CH3OH} listed in frequency for the investigated ice samples.

\subsection{Water, methanol, and \ce{CO}}
\label{wmco}
\begin{figure*}[!htp]
    \centering
    \vspace{0.5cm}
    \subfloat{\includegraphics[scale=.9]{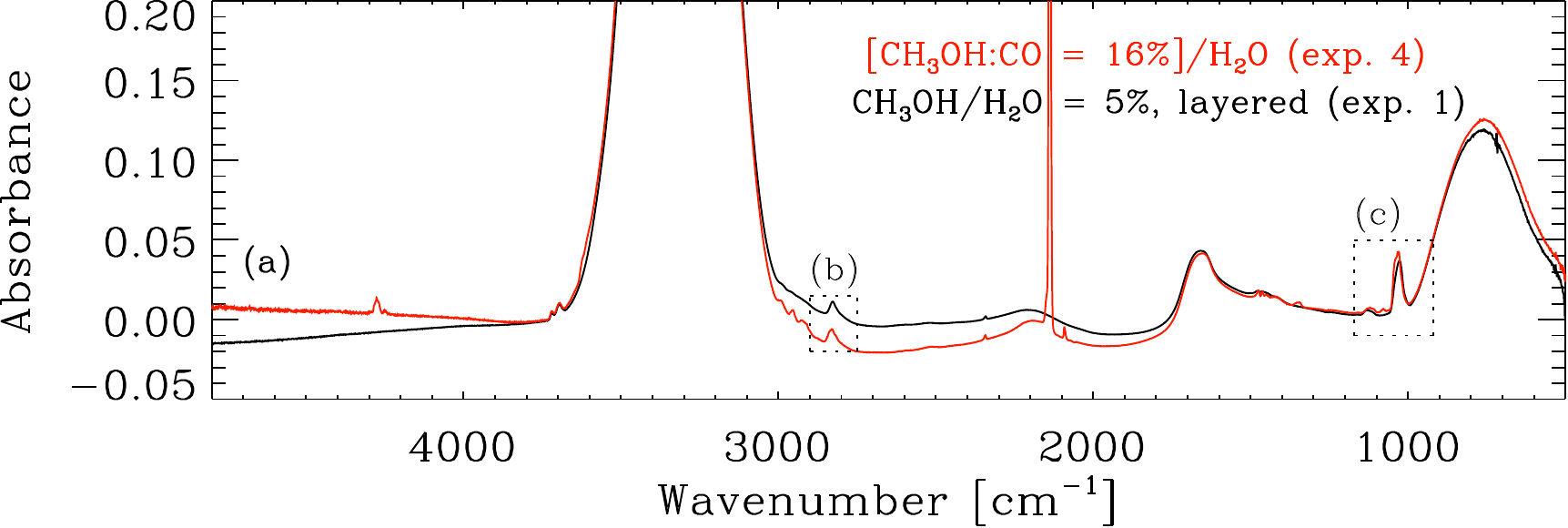}\label{CO_all}}
    
    \subfloat{\includegraphics[scale=.45]{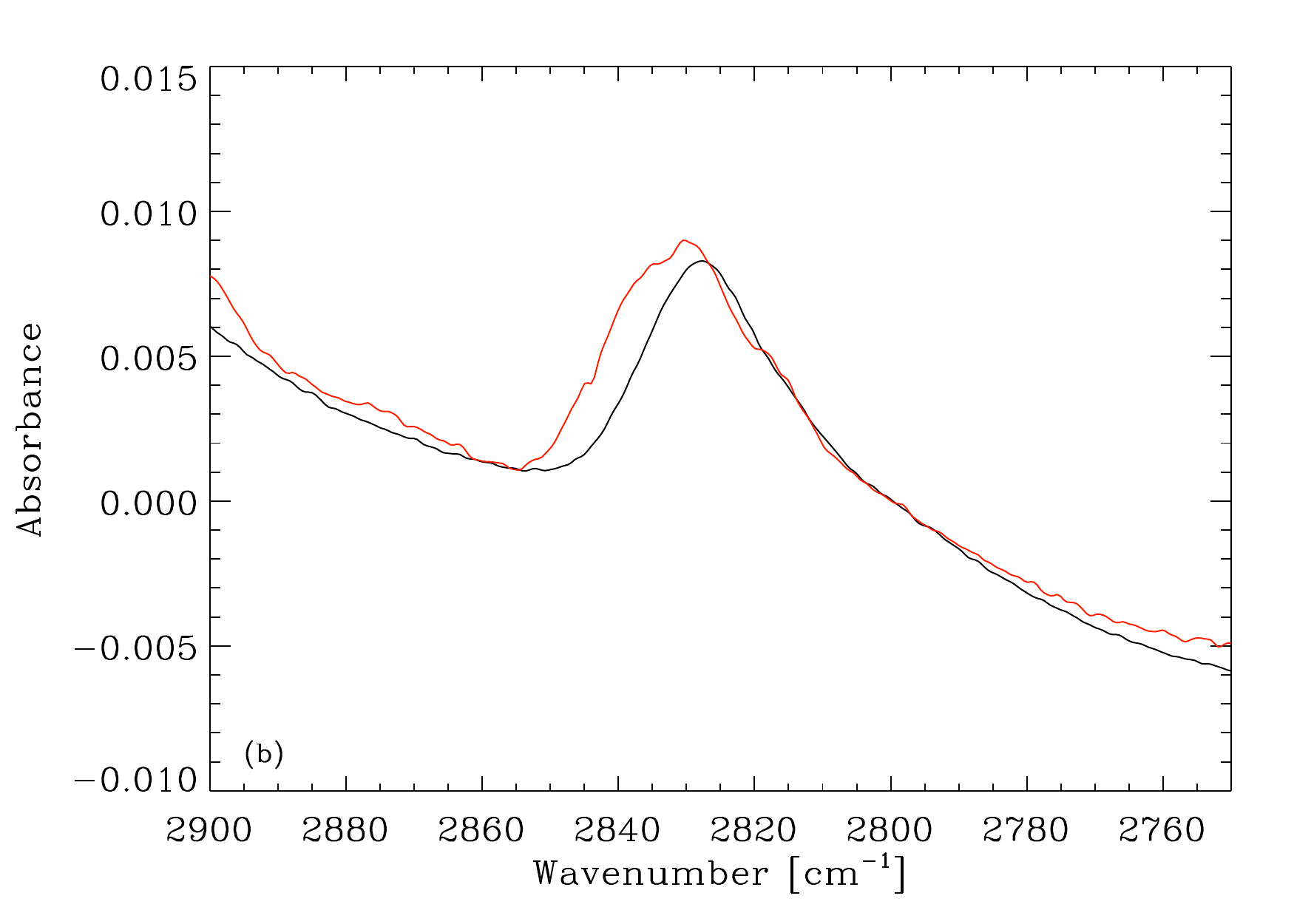}\label{CO_2828}}
    \subfloat{\includegraphics[scale=.45]{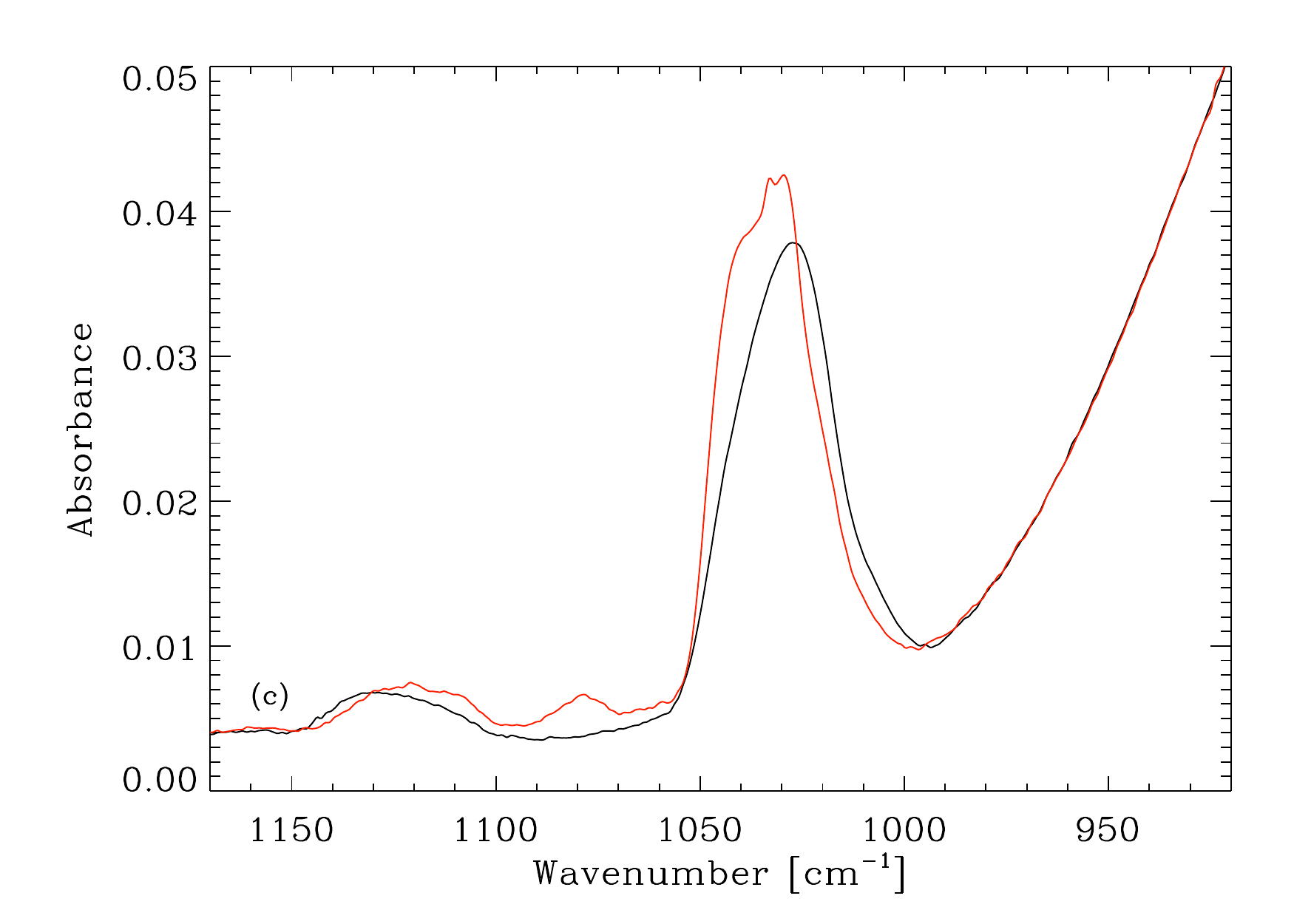}\label{CO_1032}}
    \caption{Spectra of pure methanol ice layered on top of water ice (exp. 1 in Table~\ref{exps}) compared with \ce{CH3OH}:CO ice layered on top of water ice (exp. 4 in Table~\ref{exps}) in the full frequency range (a), zoomed-in around the 3.54 $\mu$m CH-stretch band (b) and zoomed-in around the 9.75 $\mu$m CO-stretch band (c). Brackets in the labels within the panels mark species that are present in one ice layer as well as their mixture ratio. Re-scaling was applied to the spectrum of exp. 4, and in Figs.~\ref{CO_2828} and \ref{CO_1032} the spectra are shifted along the y-axis.}
    \label{CO}
\end{figure*}

\indent\indent Figure~\ref{CO} shows the comparison between the spectroscopic features of water and methanol layers (exp. 1 in Table~\ref{exps}) and layers of methanol diluted in \ce{CO} on top of the water (exp. 4 in Table~\ref{exps}).
Re-scaling was applied to the spectrum of exp. 4.
In Figs.~\ref{CO_2828} and \ref{CO_1032}, we show a zoomed-in image of the spectra around the CH- and CO-stretch features for better comparison.

As can be appreciated in these figures, the effect of diluting methanol in a CO matrix on the spectroscopic signature is different from the one observed when methanol is embedded in a water matrix. The presence of CO does not produce an appreciable shift in the frequency at the maximum of the main methanol features when compared to water. 

\begin{figure}[!htp]
    \centering
    \includegraphics[width=\hsize]{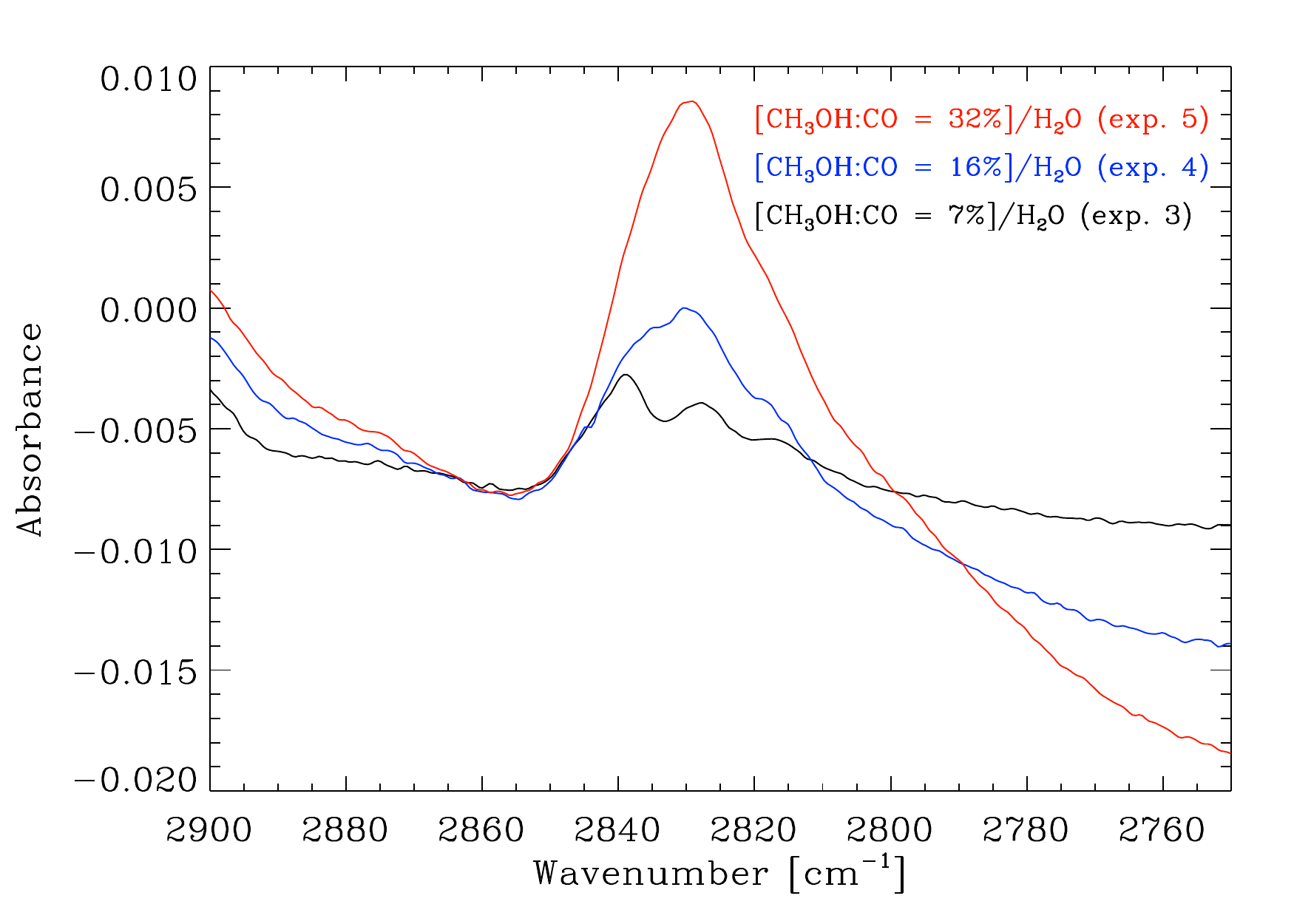}
    \includegraphics[width=\hsize]{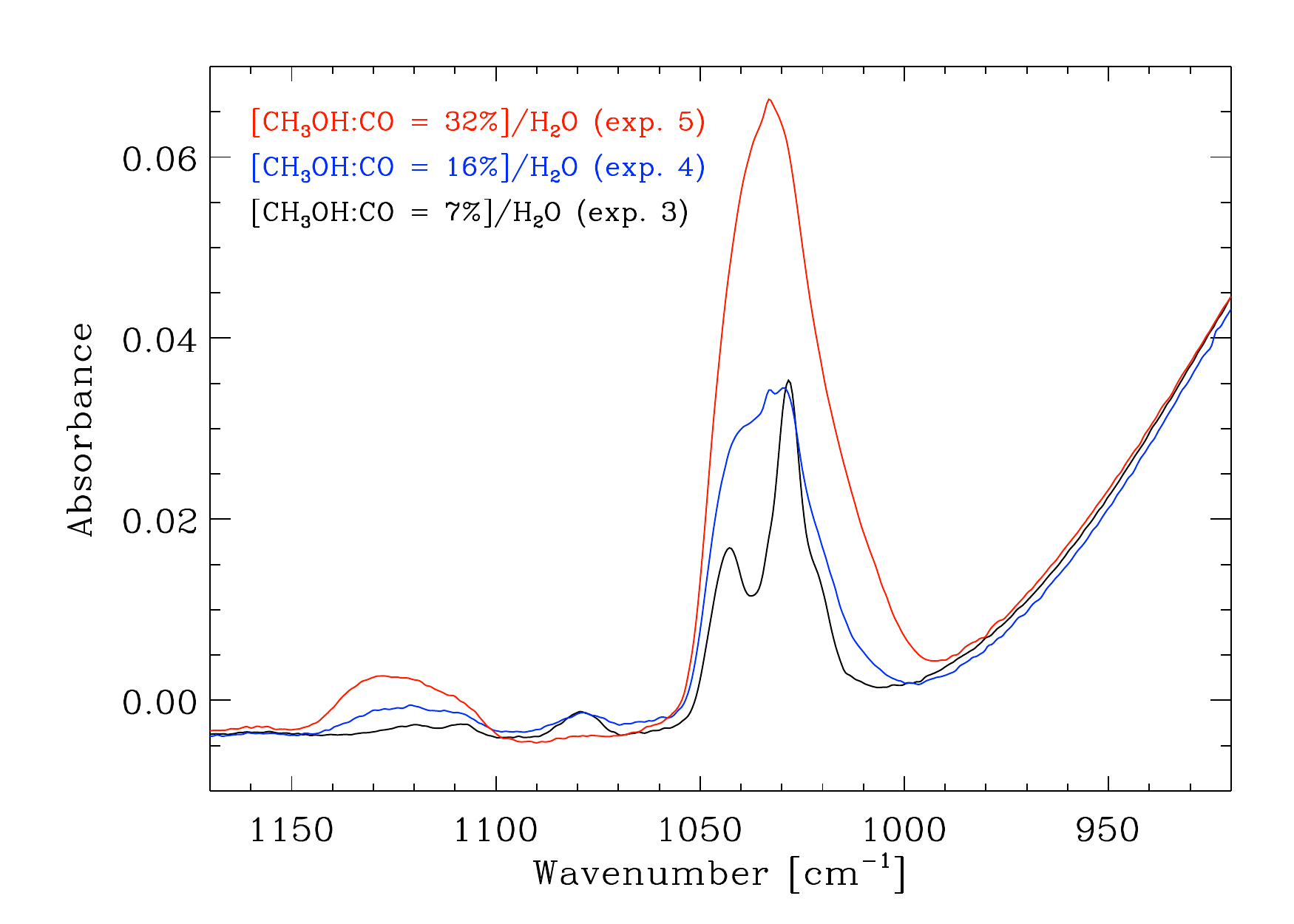}
    \caption{Spectra of methanol in CO mixture layered on top of water ice, zoomed-in on the 3.54 $\mu$m CH-stretch frequency range (top) and zoomed-in on the 9.75 $\mu$m CO-stretch frequency range (bottom). The spectra are shown for relative abundances of \ce{CH3OH}:\ce{CO} = 7\% (exp. 3 in Table~\ref{exps}), \ce{CH3OH}:\ce{CO} = 16\% (exp. 4 in Table~\ref{exps}), and \ce{CH3OH}:\ce{CO} = 31\% (exp. 5 in Table~\ref{exps}).}
    \label{COmix}
\end{figure}

Nevertheless, CO does have an effect on the profile of the absorption bands.
To further investigate the observed effect on the band shape, we prepared gas mixtures of methanol diluted in CO in different proportions (exps. 3-5 in Table~\ref{exps}).
Figure~\ref{COmix} illustrates the methanol band profile for the CH- and CO-stretch features for three different proportions of methanol in CO. 
The effect of higher dilution of methanol (around 7\%) results in a multi-component CH and CO stretch band profile that is smoothed out when the concentration of methanol is increased (from 16\% up to 32\%).
The exact position of the absorption features is listed in Table~\ref{bands}.

\subsection{Water, \ce{CO2}, \ce{CO,} and methanol}
\label{wmcoco2}
\indent\indent In this section, we present the results for mixed and layered ices with the largest number of components examined. According to \citet{2011ApJ...740..109O} and \citet{Boogert_2015}, water, carbon dioxide, carbon monoxide, and methanol are the major components of astronomical ices.
The ice growth model proposed in the paper includes the formation of layers of different molecular compositions over the molecular cloud evolution. In particular, the water ice formation is associated with the presence of solid CO$_2$, while the formation of methanol ice is believed to occur in molecular clouds through the hydrogenation of CO-rich layers that accrete on the mantles in dense cloud cores and at the edge of pre-stellar cores \citep[see also][]{2017ApJ...842...33V}.

\begin{figure*}[!htp]
    \centering
    \vspace{0.5cm}
    \subfloat{\includegraphics[scale=.9]{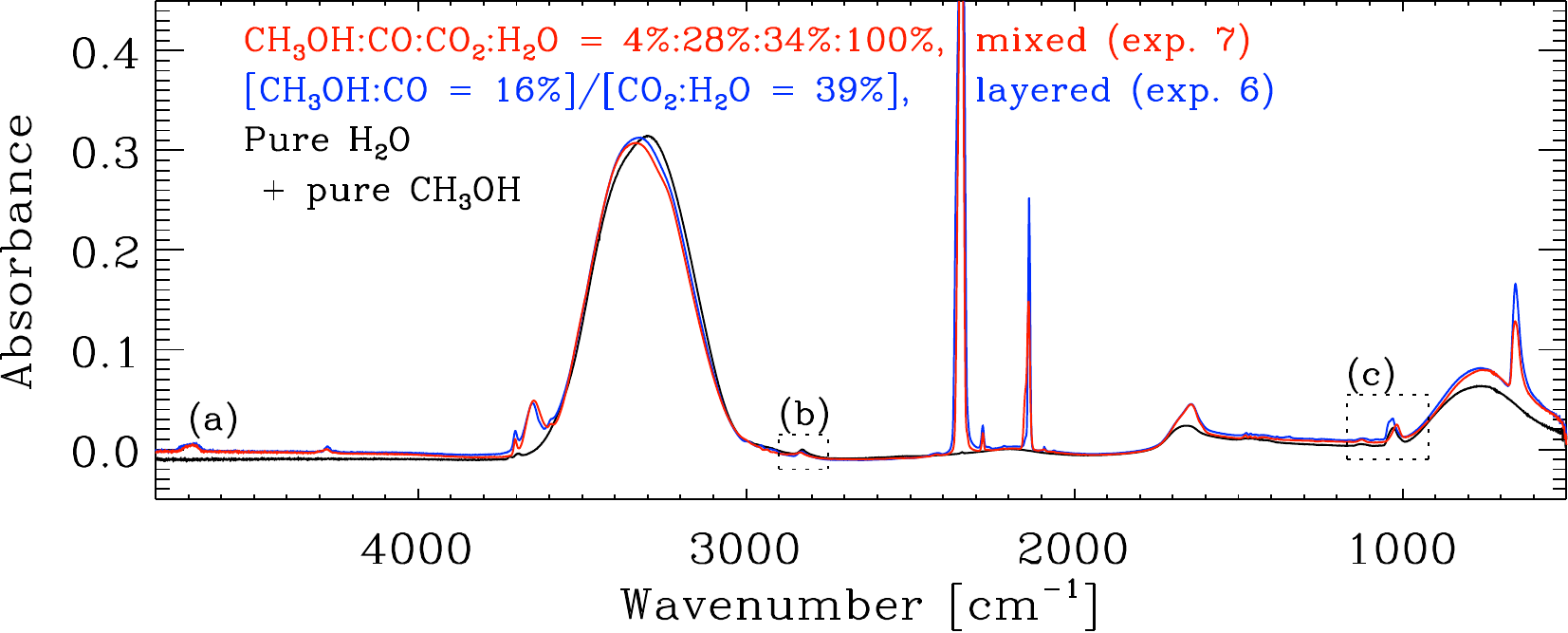}\label{CO2_all}}
    
    \subfloat{\includegraphics[scale=.45]{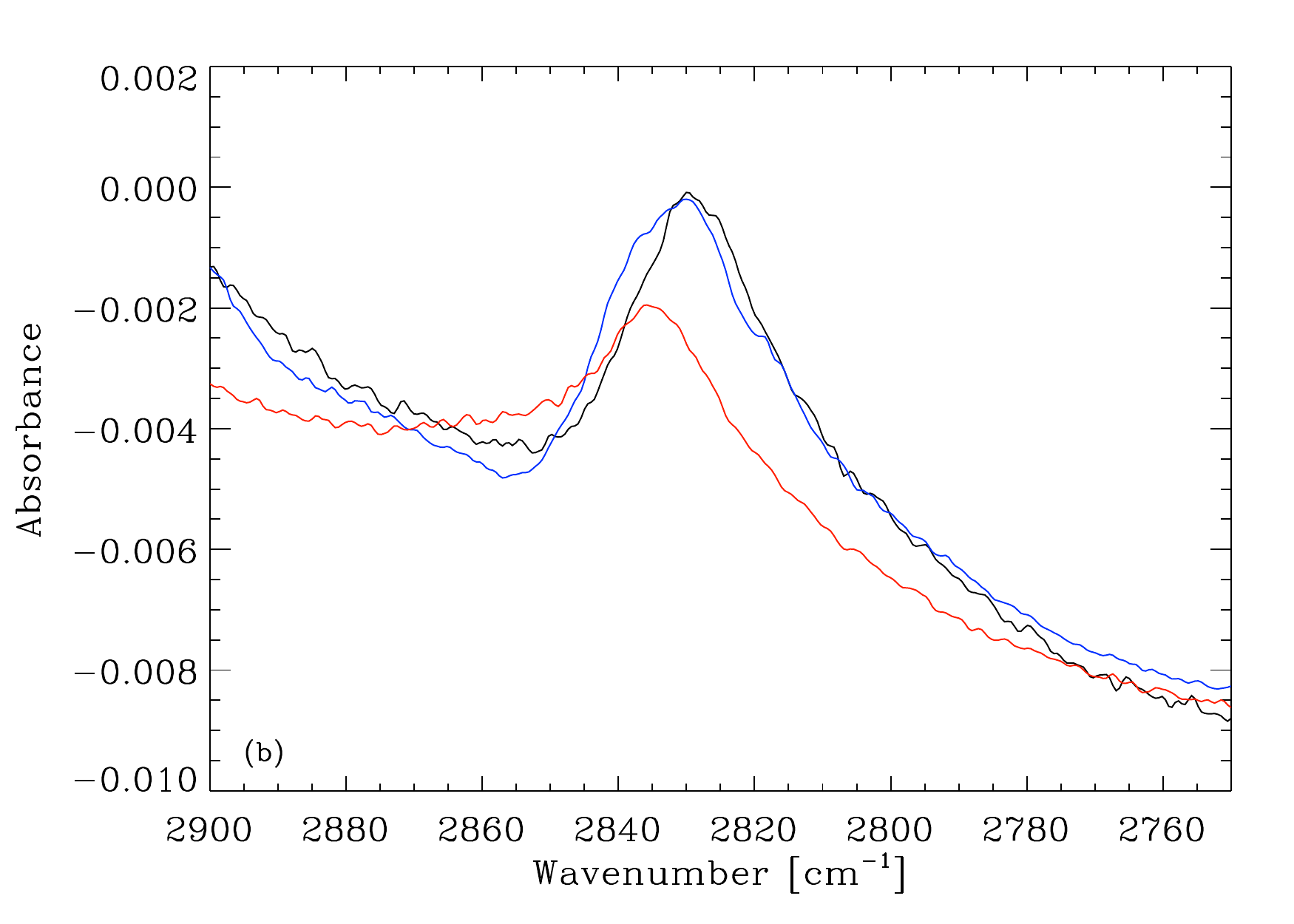}\label{CO2_2828}}
    \subfloat{\includegraphics[scale=.45]{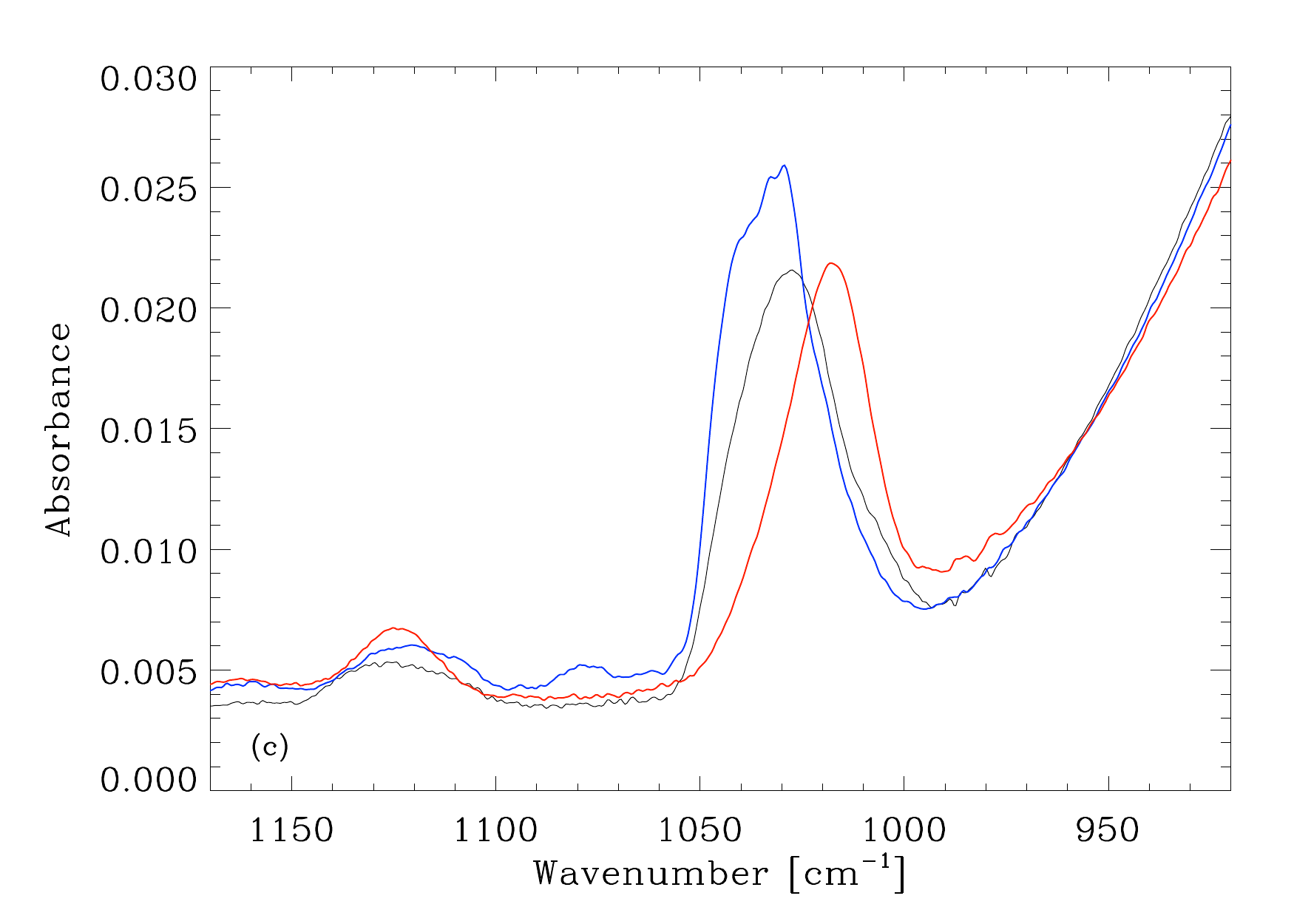}\label{CO2_1032}}
    \caption{Spectra of pure methanol and water ices compared with \ce{CH3OH}:CO ice layered on top of \ce{CO2}:\ce{H2O} ice (exp. 6 in Table~\ref{exps}) and a mixed \ce{CO2}:\ce{H2O}:\ce{CH3OH}:CO ice (exp. 7 in Table~\ref{exps}) in the full frequency range (a), zoomed-in around the 3.54 $\mu$m CH-stretch band (b) and zoomed-in around the 9.75 $\mu$m CO-stretch band (c). Re-scaling was applied to the spectrum of exp. 7, and in Figs.~\ref{CO2_2828} and \ref{CO2_1032} the spectra are shifted along the y-axis.}
    \label{CO2}
\end{figure*}

Therefore, two different ice mixtures were considered to simulate this ice structure in the laboratory, following the molecular ratio in \citet{2011ApJ...740..109O}. The first mixture (exp. 6 in Table \ref{exps}) includes \ce{CO2} in water at about 40\% molecular abundance, whereas the second one consists of methanol diluted in CO at a ratio of about 15\%. The two mixtures were deposited in layers, starting with the water-rich one and increasing the thickness of the CO-rich mixture until a relative proportion of approximately 30\% between CO and water was reached. The recorded spectra are compared with an ice layer formed by condensation of a four components gas mixture (exp. 7 in Table \ref{exps}) in equal relative proportion to that in exp. 6 (see Table \ref{exps}).

Figure~\ref{CO2} illustrates the corresponding spectroscopic data, in the full experimental frequency range, and shows a zoomed-in view around the most prominent methanol features.
Re-scaling was applied to the spectrum of the mixed \ce{CH3OH}-CO-\ce{CO2}-\ce{H2O} ice.
Also, the comparison with the summed spectra of pure water and methanol is shown, corrected for the total ice thickness.
We see clear shifts in the maximum band position of the CH-stretch band as well as the CO-stretch band for mixed and layered ices. The shape of the CO-stretch band for the layered ice shows similarities with exp. 4 in Table~\ref{exps} (cf. Fig.~\ref{COmix}). Since only the composition of the bottom layer is different for exps. 4 and 6 in Table~\ref{exps}, this hints at the assumption that there is no strong interaction between the species in the upper and lower layers.

\subsection{Warm-up}
\label{warming}
\indent\indent In our experiments, we aim to characterise the change in the methanol spectroscopic features observed during ice warm-up, which is relevant for protostellar cores in the neighbourhood of the young stellar object. The layered and mixed ice samples were heated up to 150~K, the temperature at which the methanol ice desorbs and the water ice transitions to a crystalline state.
The crystallisation can be accompanied by desorption of water ice following a desorption rate profile peaking around 180 K \citep[cf.][]{Potapov_2018}. Similar desorption temperatures were observed by \citet{Collings_2004}, who measured desorption temperatures around 160 K for a heating rate of 0.08 K s$^{-1}$.

\begin{figure*}[!htp]
    \centering
    \vspace{0.5cm}
    \subfloat{\includegraphics[scale=.5]{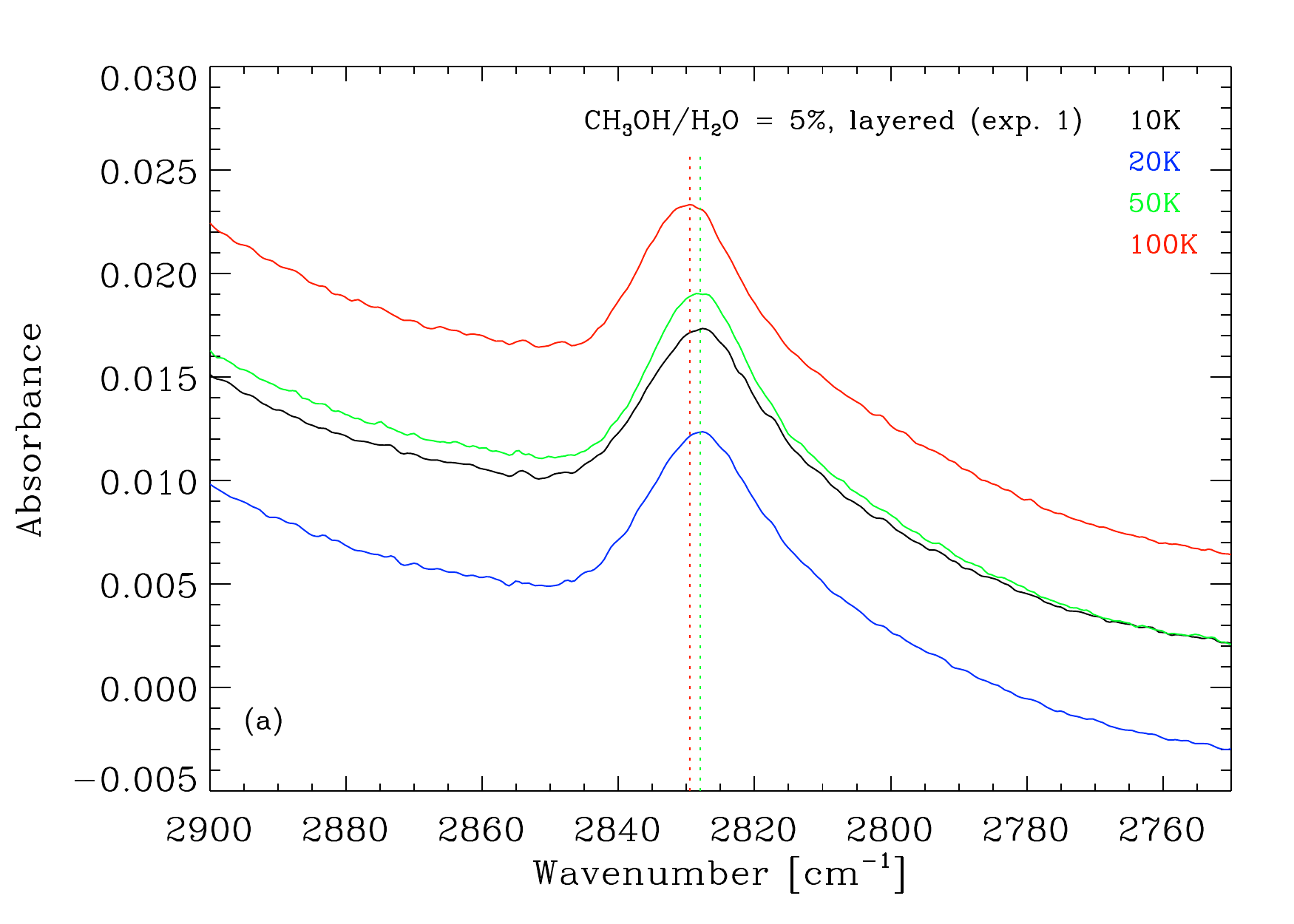}\label{warm_2comp_layer_2828}}
    \subfloat{\includegraphics[scale=.5]{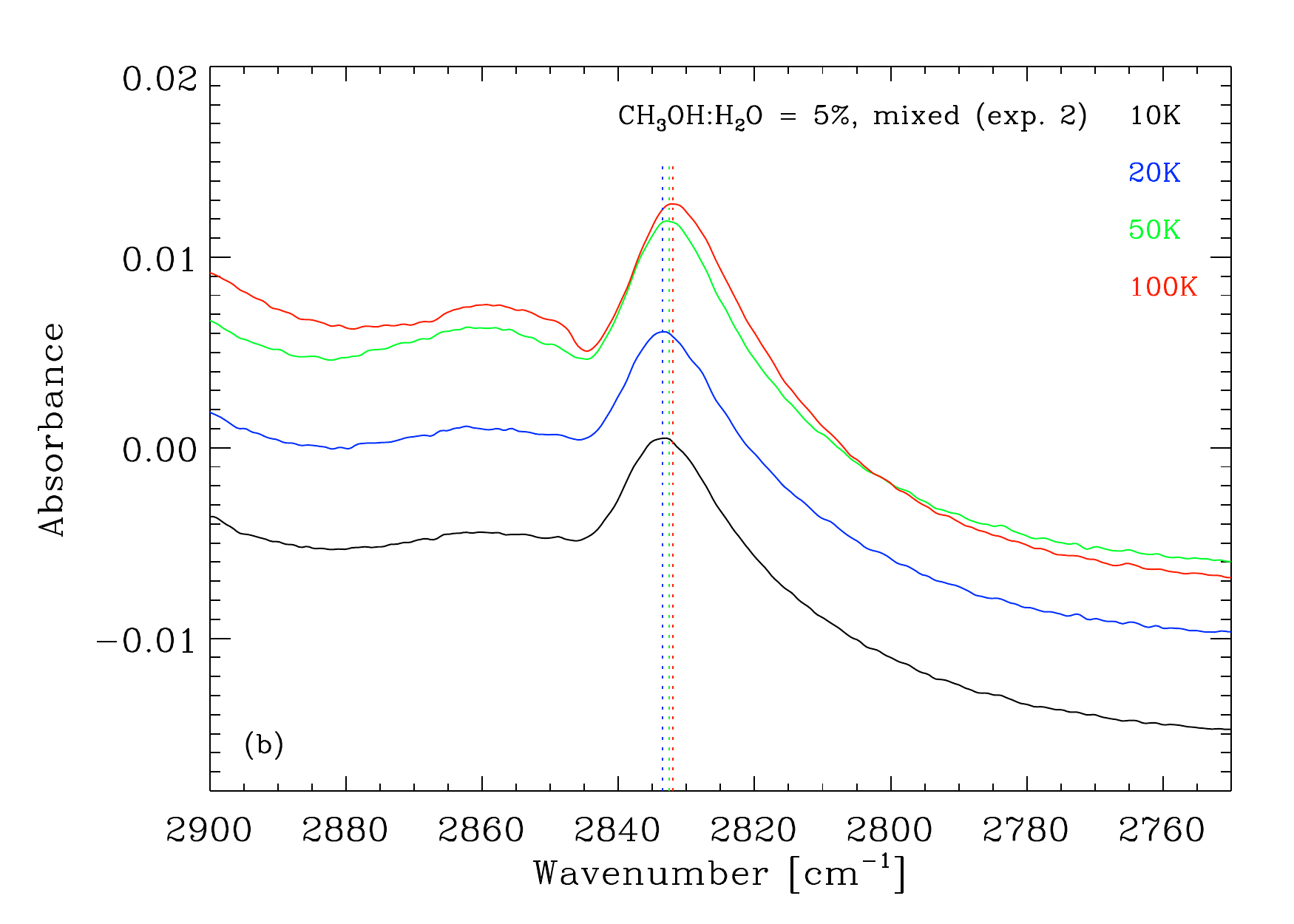}\label{warm_2comp_mix_2828}}
    
    \subfloat{\includegraphics[scale=.5]{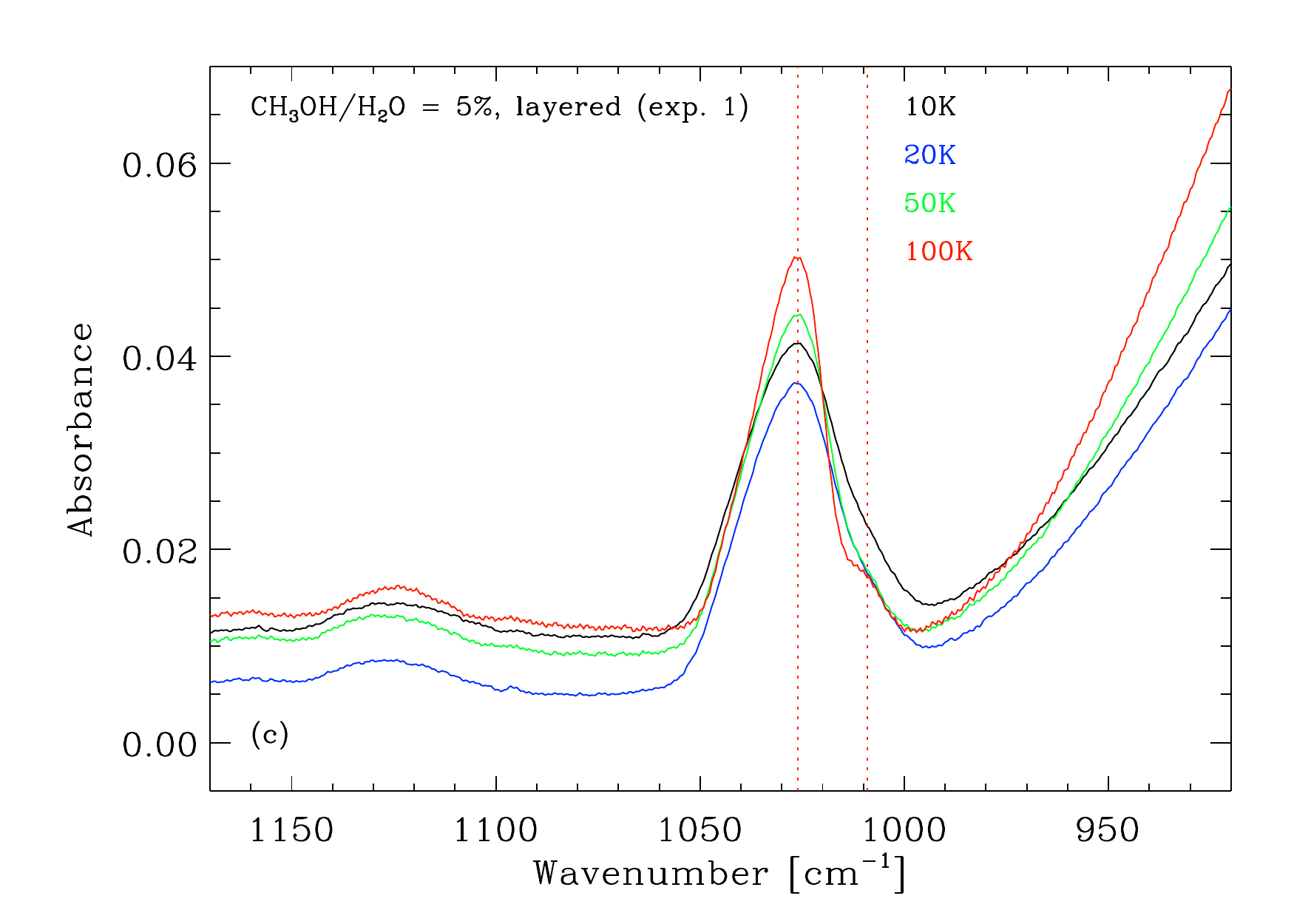}\label{warm_2comp_layer_1032}}
    \subfloat{\includegraphics[scale=.5]{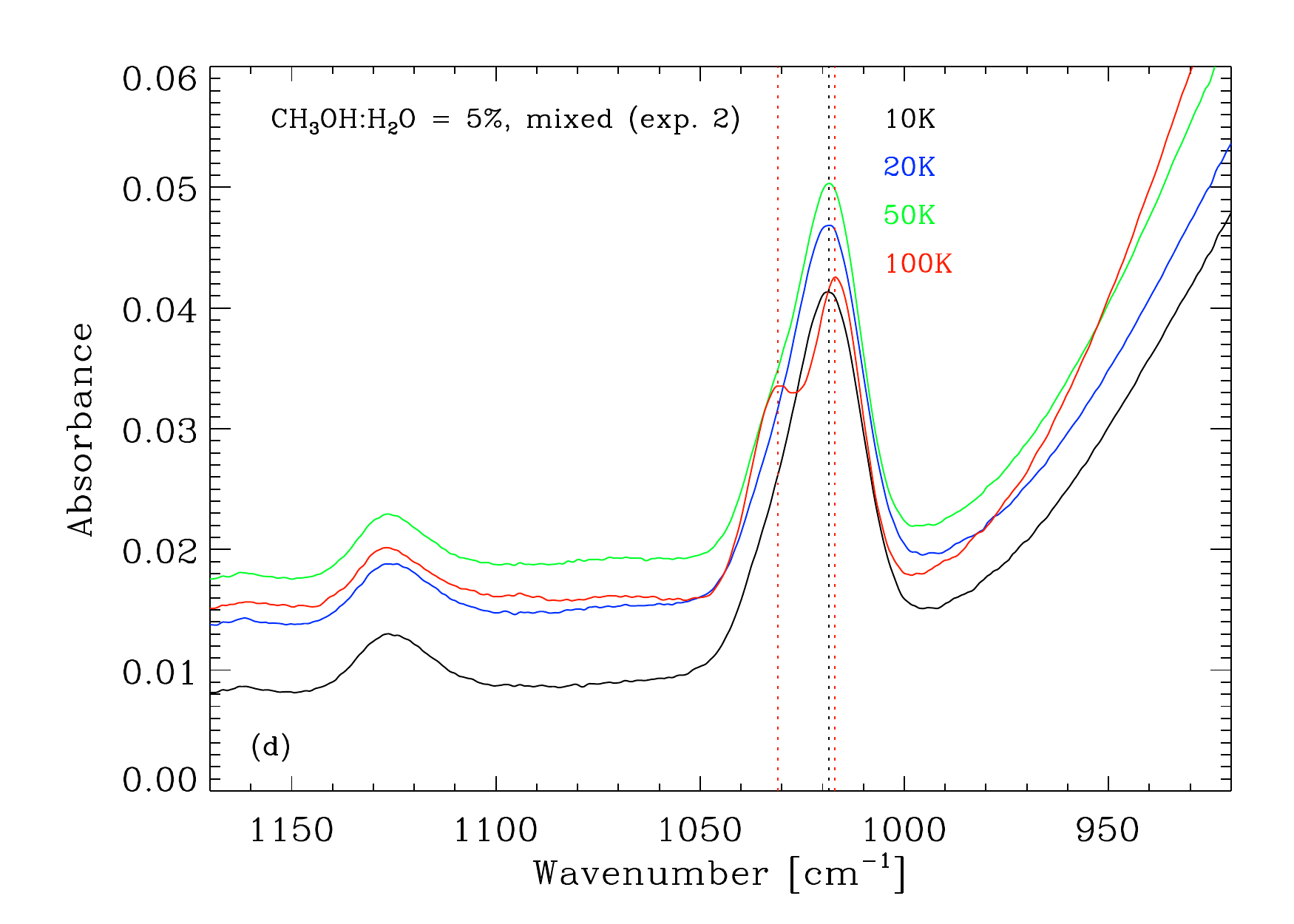}\label{warm_2comp_mix_1032}}
    \caption{Heating of methanol layered on top of water ice (exp. 1 in Table~\ref{exps}) and mixed \ce{CH3OH}:\ce{H2O} ice (exp. 2 in Table~\ref{exps}), zoomed-in on the 3.54 $\mu$m CH-stretch band, layered (a) and mixed (b) ices, and zoomed-in on 9.75 $\mu$m CO-stretch band, layered (c) and mixed (d) ices. For heated samples, vertical coloured lines mark the position of local band maxima and shoulder features for the respective temperature.}
    \label{warm_2comp}
\end{figure*}

The first result we show in Fig.~\ref{warm_2comp} is the effect of the heating of methanol layered on top of water ice as well as a mixed \ce{H2O}:\ce{CH3OH} in direct comparison. Comparing with \citet{Dawes2016}, Fig.~\ref{warm_2comp} shows that upon heating to 100 K the layered \ce{H2O}/\ce{CH3OH} ice is mixed, while the initially mixed \ce{H2O}:\ce{CH3OH} ice shows segregation of methanol.

\begin{figure*}[!htp]
    \centering
    \vspace{0.5cm}
    \subfloat{\includegraphics[scale=.5]{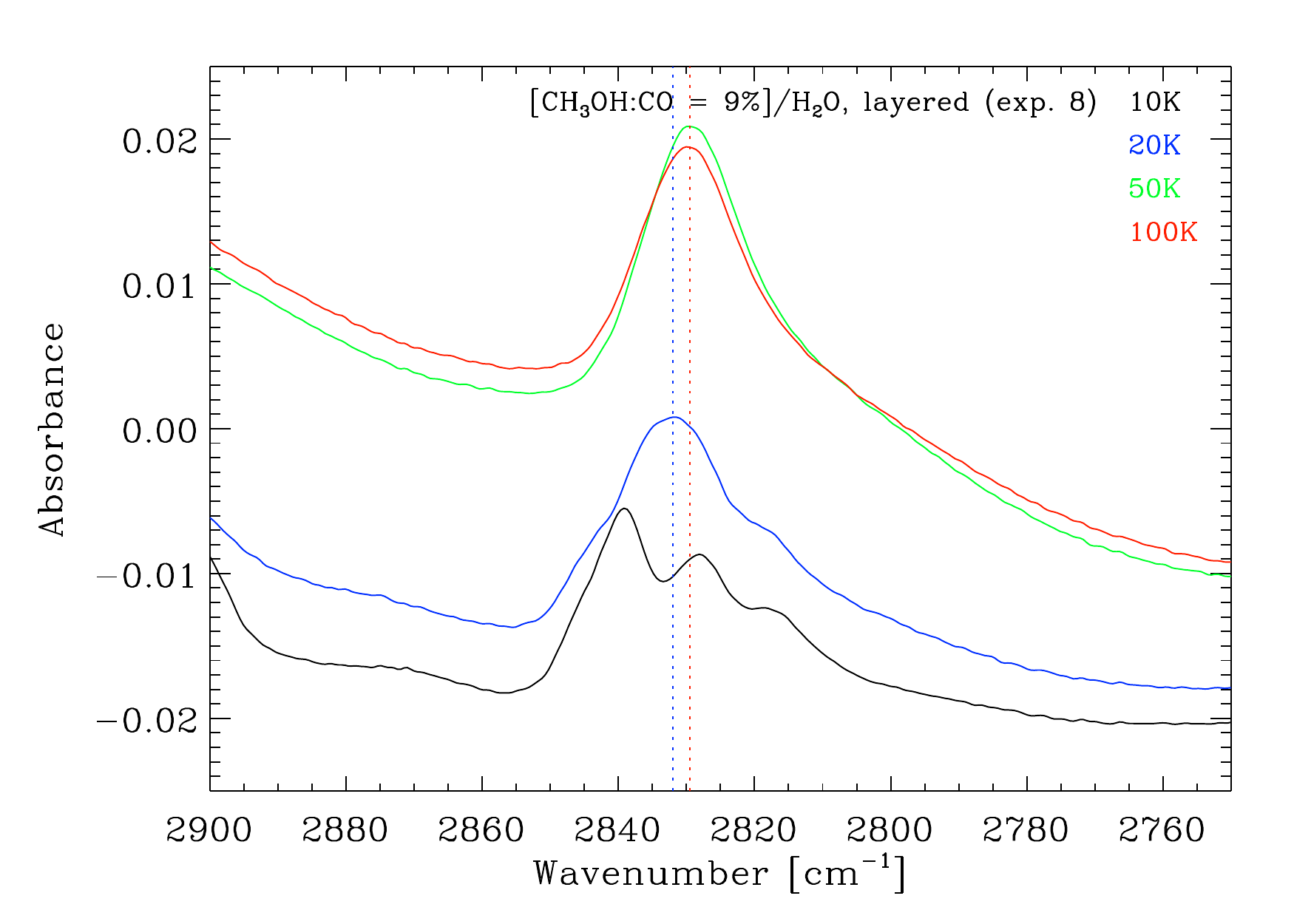}}
    \subfloat{\includegraphics[scale=.5]{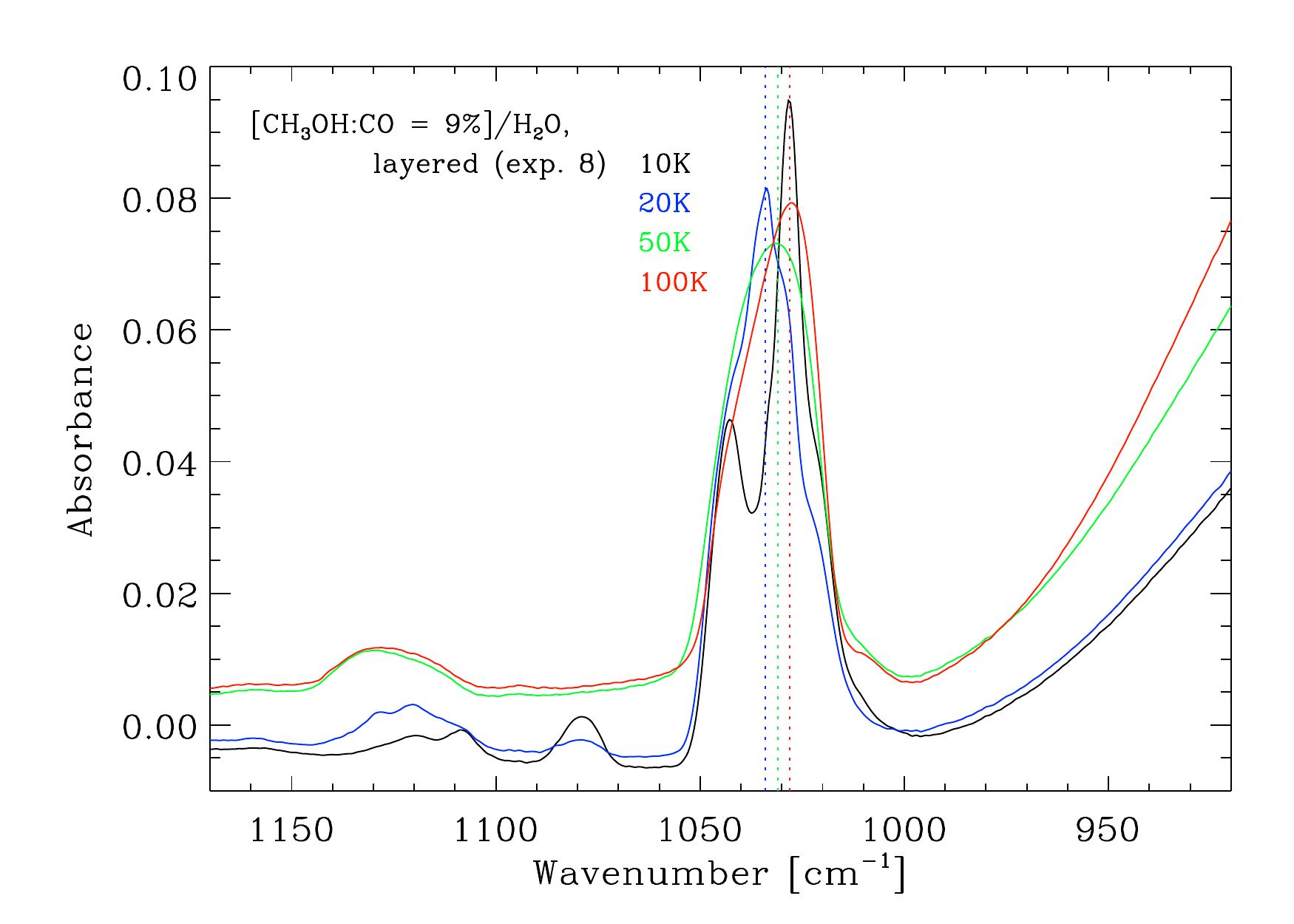}}
    \caption{Heating of \ce{CH3OH}:CO/\ce{H2O} ices, zoomed-in on the 3.54 $\mu$m CH-stretch band (left) and 9.75 $\mu$m CO-stretch band (right). For heated samples, vertical coloured lines mark the position of the band maximum for the respective temperature.}
    \label{CO_warm}
\end{figure*}

Secondly, methanol diluted in CO was layered on top of water ice. For this particular experiment (see exp. 8 in Table~\ref{exps}), the proportion between water and methanol was increased to 10\% as this is the methanol-to-water ratio constrained by \citet{Goto_2020} after comparing their observations with laboratory work by \citet{1993ApJS...86..713H}.
The increase of the methanol-to-water ratio is followed by a consequent change in the \ce{H2O}:CO ratio to 113\%, and methanol to CO ratio to 9\%. In Fig.~\ref{CO_warm}, the spectra recorded at 10, 20, 50, and 100~K are shown.
For 10 K, we see no effect on the CH-stretch band or the CO-stretch band compared to exp. 3 in Table~\ref{exps}, which has a similar \ce{CH3OH}:CO ratio to exp. 8 in Table~\ref{exps}.
The effect of the heating on the methanol band profile is quite significant, starting with very mild heating already at 20~K presenting a spectroscopic signature compatible with methanol segregation in the CO matrix.

\begin{figure*}[!htp]
    \centering
    \subfloat{\includegraphics[scale=.5]{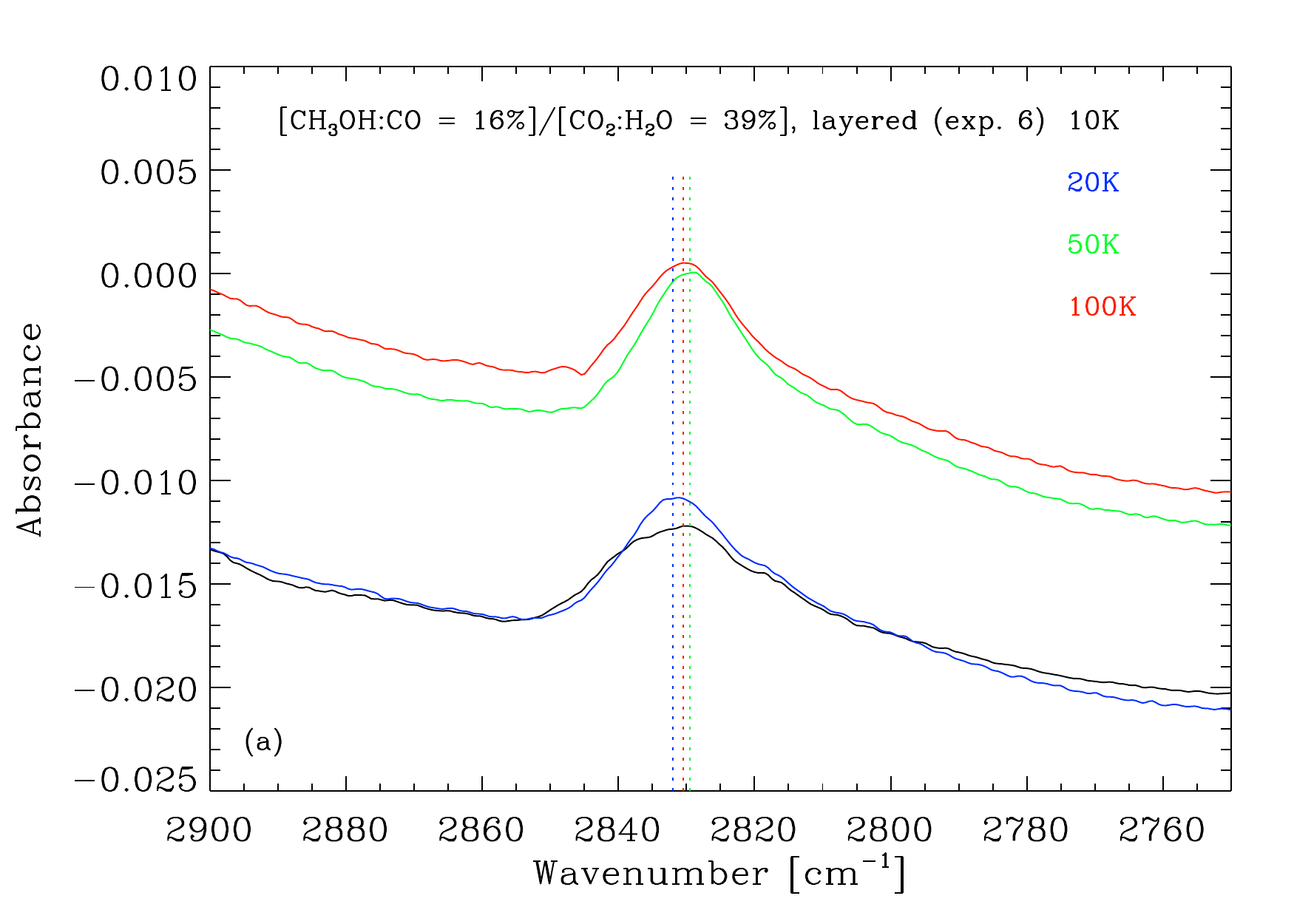}\label{warm_4comp_layer_2828}}
    \subfloat{\includegraphics[scale=.5]{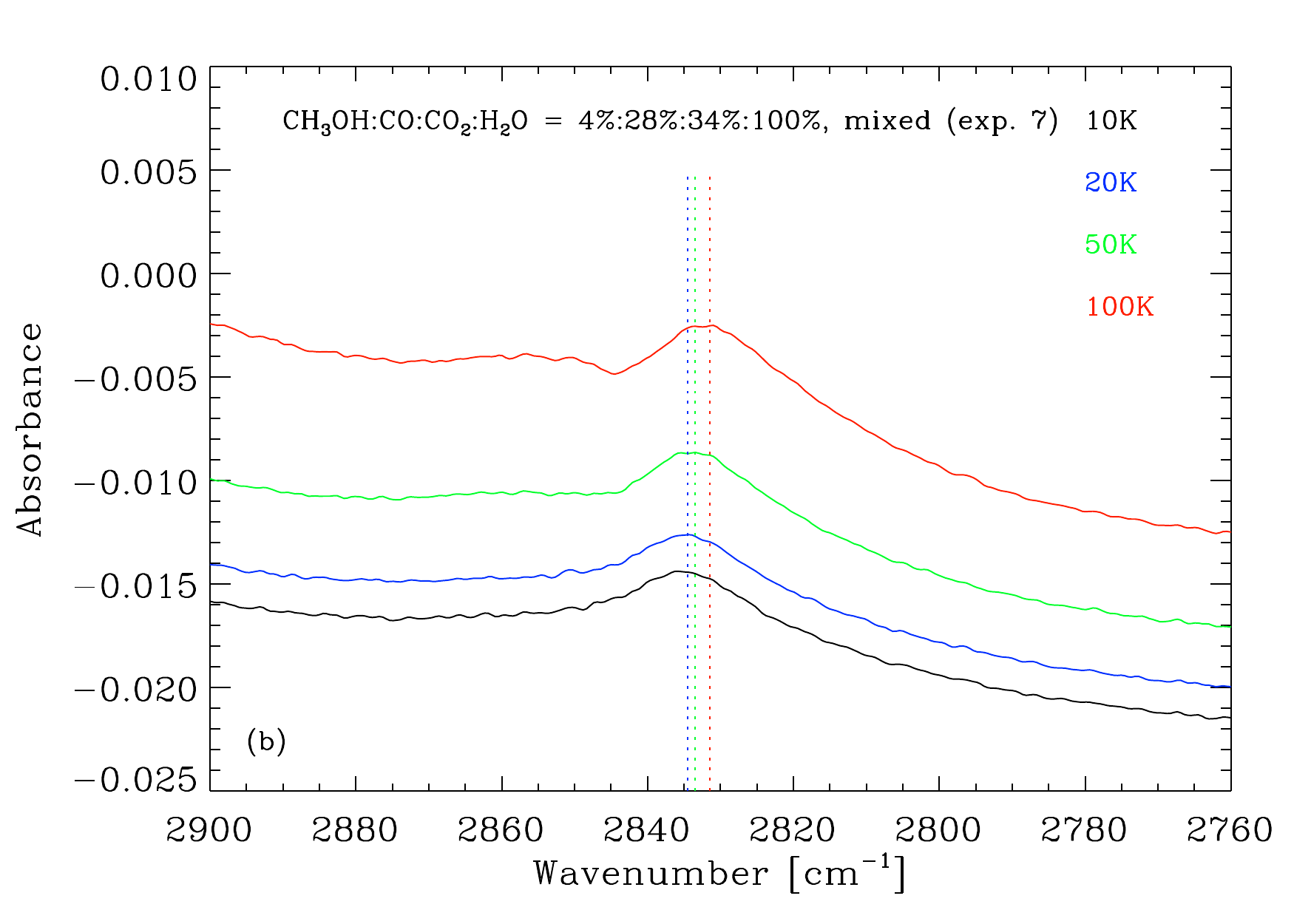}\label{warm_4comp_mix_2828}}
    
    \subfloat{\includegraphics[scale=.5]{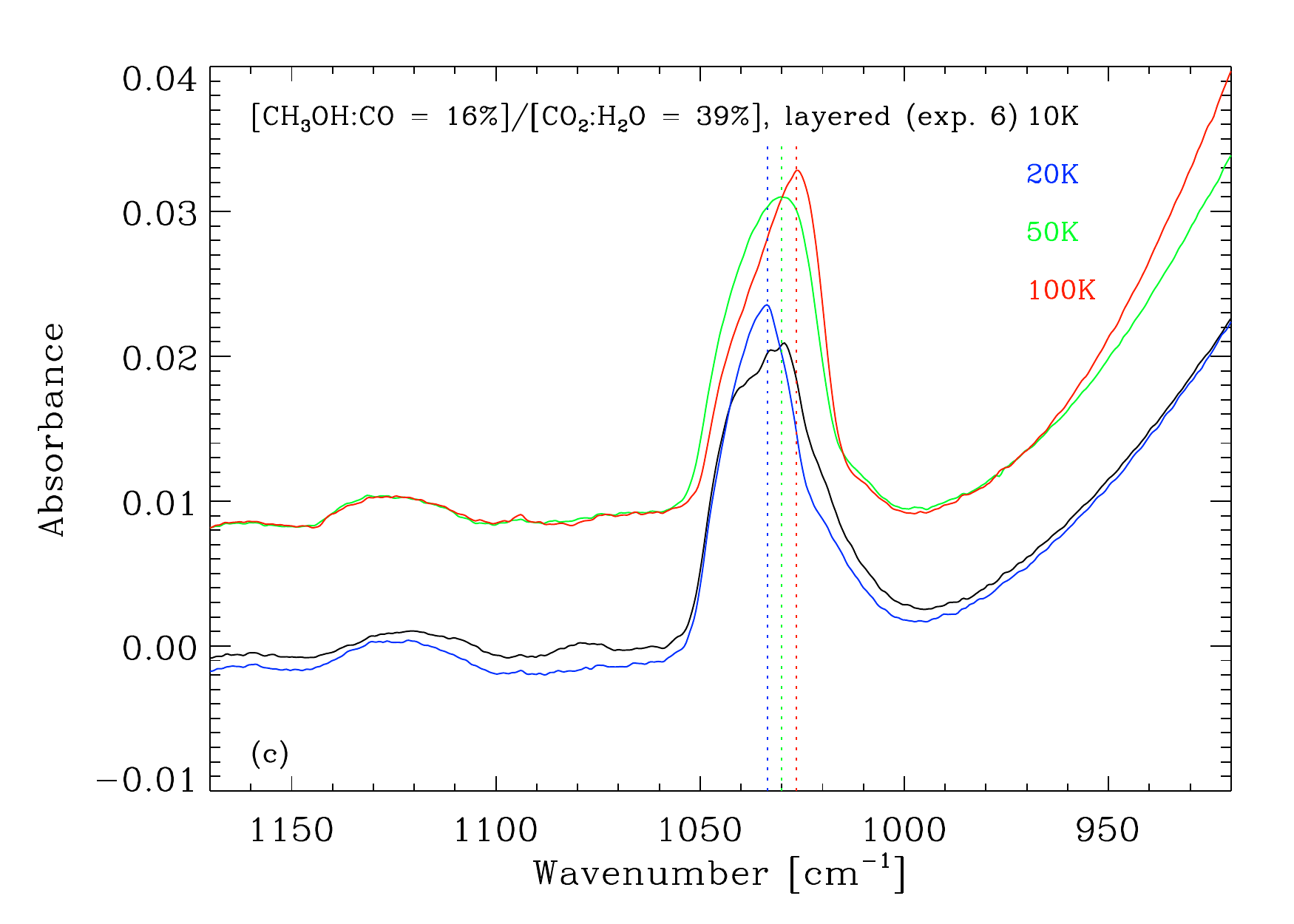}\label{warm_4comp_layer_1032}}
    \subfloat{\includegraphics[scale=.5]{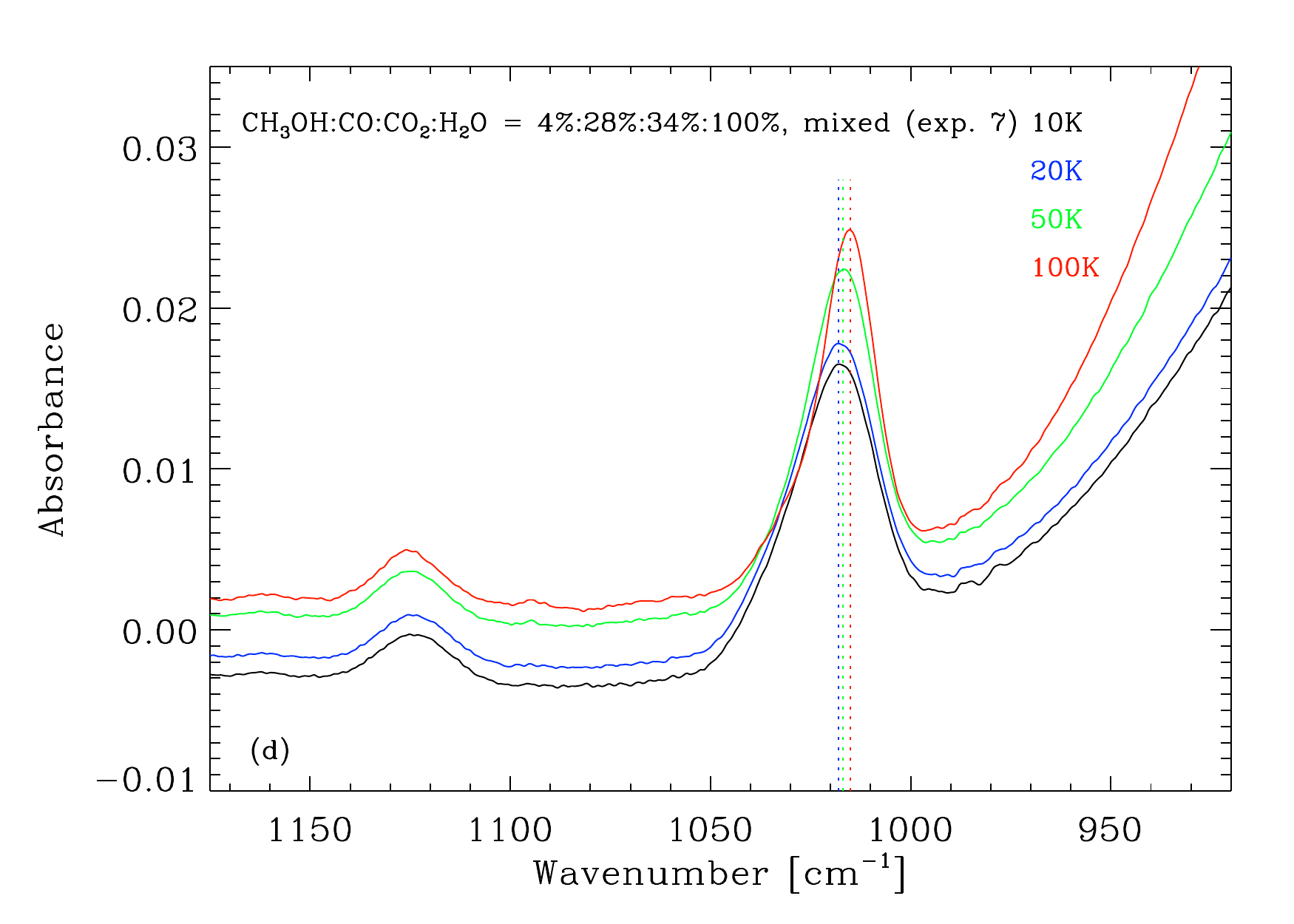}\label{warm_4comp_mix_1032}}
    \caption{Heating of \ce{CH3OH}:CO layered on top of \ce{CO2}:\ce{H2O} ice (exp. 6 in Table~\ref{exps}) and mixed \ce{CH3OH}:CO:\ce{CO2}:\ce{H2O} ice (exp. 7 in Table~\ref{exps}), zoomed-in on the 3.54 $\mu$m CH-stretch band, layered (a) and mixed (b) ices, and zoomed in at 9.75 $\mu$m CO-stretch band, layered (c) and mixed (d) ices. For heated samples, vertical coloured lines mark the position of the band maximum for the respective temperature.}
    \label{warm_4comp}
\end{figure*}

The results of our measurements for the more complex \ce{H2O}:\ce{CO2}:CO:\ce{CH3OH} ice samples, up to 100 K, are shown in Fig.~\ref{warm_4comp}.
The 3.54 $\mu$m (2828 cm$^{-1}$) and 9.75 $\mu$m (1026 cm$^{-1}$) methanol bands show different profiles when methanol is mixed in a water matrix compared to a layered structure.

In summary, for mixed and layered \ce{H2O}-\ce{CH3OH} ices (exps. 1 and 2 in Table~\ref{exps}) we are not able to deduce the ice temperature from the band shape and position for T < 100~K, where segregation or mixing begins.
This changes for ices with an increased number of species where not only the presented \ce{CH3OH} band shapes but also the band positions are strongly affected by the temperature of the ice, even at T = 20~K (cf. exps. 6-8 in Table~\ref{exps}). Since these ice analogues are more similar to the expected ice composition of interstellar ices, the presented measurements can help to put constraints on the temperature of observed ices.

\subsection{Inverse deposition}
\label{inv}
\indent\indent Following the approach presented by \citet{2007A&A...472..691G}, \citet{2008JPCA..112..457M}, and \citet{C3CP53767F}, among other studies, we investigated the effect of inverse deposition (ID) of ice layers on their spectroscopic signature. We limited this study to the \ce{H2O}:\ce{CH3OH} samples, depositing a layer of water ice on top of a layer of methanol ice, keeping the relative proportion of methanol with respect to water to 5\%.
We then compared the spectra of the inverse deposited ice sample with the data from previous sequential deposition (exp. 1, Table~\ref{exps}), observing that there is no appreciable effect on the shape and position of the absorption bands due to the deposition order of the two components.
We warmed up the ice samples to 150 K, in steps of 20, 50, and 100 K, and we find no difference in the spectroscopic features between sequential and inverse sequential deposition due to this heating.

\section{Discussion and comparison with previous work}
\label{disc}
\indent\indent The spectroscopic features of the considered ices show some general trends when we compare the spectra recorded in layers with the ones for completely mixed ices. 
The chemical environment of a molecular species embedded in the ice can affect its spectral band shape due to the effect of the intermolecular interactions.
It is therefore important to measure the changes in band features of abundant ice components, such as methanol ice, so that future sensitive observations can provide information about the ice structure and composition.
As a result of the different interactions, we observe shifts in frequency, as well as changes in the band profile.

Intuitively, it is expected that the spectral features of molecular species deposited in layers are more similar to the spectra of the corresponding pure ices than the mixed equivalent. 
When methanol is premixed with other molecular species in the gas phase, its band shapes in the solid phase will diverge from those of the pure ice. This effect is caused by the interaction between the different species, which has an effect on the vibration of the molecular bonds and has been studied by various groups such as \citet{1990ApJ...355..357S} and \citet{1999A&A...350..240E}.
In contrast, when methanol is deposited as a layer on top of other species, the shape of the absorption bands stays close to that of the pure methanol since the molecular vibrations are more similar to those of pure ices due to the missing influence of other species.

This problem has been addressed in previous works \citep{2007A&A...472..691G,2010PCCP...12.3164H,2009ApJ...690..486M}, in which the interaction of binary ice mixtures, sequentially or co-deposited, has been investigated. 
The molecular species considered in these studies were \ce{H2O}:\ce{CO2}, \ce{CO2}:\ce{CH3OH,} and \ce{H2O}:\ce{CH4}. The analysis focussed on adsorption and diffusion processes between the investigated species, and it provided information on the spectral changes observed. These changes have been interpreted as variations in terms of the structure that the host species can adopt within the guest one. 
The results focussed on the \ce{CO2} or \ce{CH4} features as a probe for the change in the water or methanol environment. The authors conclude that the distortions of the molecular environment must be caused by weak interactions between the host and the guest species, where the porosity of the host matrix plays the main role in setting the guest absorption properties.

\citet{1990ApJ...355..357S} reported an investigation of \ce{CO2} co-deposited with samples of different compositions and concentrations, including methanol. 
Similarly to our results, they observed significant variations in position and width of the \ce{CO2} vibrational bands, which are associated with the relative concentration of the two species, as well as by heating from 10 to 150 K.

In recent observations of the L1544 starless core at 3 $\mu$m by \citet{Goto_2020}, laboratory data from \citet{1993ApJS...86..713H} have been used to constrain the methanol to water ice ratio. These data were recorded for an ice mixture that has been co-deposited from the gas phase, with the molecular composition of \ce{H2O}:\ce{CH3OH}:CO:\ce{NH3} = 100:10:1:1, which might not be representative of the average ice composition in cold clouds. 
In particular, the effect of \ce{CO2} and CO on water-rich ices was not taken into account, nor was the role of layered versus co-deposited ice samples on the spectral features.

The present work shows a new approach to the understanding of the ice structure as the experiments are based on compositions that resemble interstellar ices. We focus our attention on \ce{CH3OH}, as this was recently found to be quite abundant in the direction of the pre-stellar core L1544 \citep{Goto_2020}.
The main points addressed consider the effect of a layered structure versus a mixed one, how the increase of the number of components impacts the band peak frequency and shape, and the effect of heating on the spectroscopic changes.

In all the experiments performed, the layered ice structure has significantly changed the absorption features compared to the ice samples deposited from pre-mixed gases. 
The ice layers present strong similarities with the spectra from pure ice samples, while in mixed ices the frequency of the absorption bands can be shifted from those assigned to pure ices. In this case, the polarity of the major ice component, which acts as hosting matrix, is probably the main thing responsible for the spectroscopic changes.
Table~\ref{bands} shows the change in the absorption frequency positions in the different experiments.
Our results, when compared with the observational work of \citet{Goto_2020} and \citet{Dawes2016}, favour a layered structure of interstellar ices.

\begin{table}[ht]
\centering
\caption{Peak wave number (cm$^{-1}$) of the 3.54 and 9.75 $\mu$m methanol bands in the experiments presented in the paper, recorded at 10 K. Multiple band assignments connect to several local maxima or shoulder features for one band. The error in the measured positions lies within 2 cm$^{-1}$.}
\label{bands}
\begin{tabular}{l  c  c}
\hline \hline
Experiment$^{a}$ & \multicolumn{2}{c}{Wave number (cm$^{-1}$)} \\ [0.5ex]
& $\nu_3$ CH stretch$^{b}$ & $\nu_8$ CO stretch$^{b}$ \\
\hline \\ [0.15ex]
Pure \ce{CH3OH} & 2829           & 1028 \\
1               & 2828           & 1028 \\
2               & 2833           & 1018 \\
3               & 2839;2828;2817 & 1043;1028;1021 \\ 
4               & 2830           & 1040;1033;1029 \\ 
5               & 2830           & 1033 \\ 
6               & 2832           & 1041;1038;1033;1029 \\ 
7               & 2834           & 1018 \\ 
8               & 2839;2828;2817 & 1043;1028;1021 \\ 
9               & 2828           & 1030 \\ [0.5ex]
\hline
\end{tabular}

\begin{list}{}
\item (a) See Table~\ref{exps}.
\item (b) The vibrational assignment is taken from \citet{1993ApJ...417..815S}, where $\nu_3$ and $\nu_8$ are assigned to 2827 cm$^{-1}$ and 1026 cm$^{-1}$, respectively.
\end{list}
\end{table}

By comparing the spectra of methanol diluted in water or the CO matrix, we see that the band shifts are significantly different (cf. Figs.~\ref{CH3OH}, \ref{CO}, and \ref{CO2} relating to exps. 2, 4, 6, and 7 in Table~\ref{exps}). The water matrix induces the largest shift, while no significant changes are observed in CO. 
Considering the very polar nature of methanol, which is prone to form hydrogen bonds, we could conclude that the intermolecular interactions with water are stronger than with CO, and this effect is reflected in the spectroscopic signature.

\citet{Dawes2016} compared \ce{CH3OH} band positions of pure \ce{CH3OH}, pure \ce{H2O,} and mixed \ce{CH3OH}-\ce{H2O} ices. They observed a blueshift of the C-H stretching band for mixed ices compared to pure \ce{CH3OH,} and they describe a red shoulder feature for the C-O stretching component that gets more dominant as methanol is more diluted in water. The described band positions agree well with our measurements of mixed \ce{H2O} and \ce{CH3OH} ices (exp. 2 in Table~\ref{exps}).

The degree of dilution of the minor ice component  also has an effect on the band's appearance. The effect observed for three different methanol abundances in CO shows quite different band shapes, especially for the most diluted one.
We could relate this effect to a major isolation of methanol in the CO environment, bringing it to different binding sites, while for the most concentrated mixtures the methanol can start to segregate, resembling more closely the features of the pure ice.

The heating of the most diluted \ce{CH3OH}:\ce{CO} mixtures shows interesting results (Fig.~\ref{CO_warm}). 
At a temperature of 20 K, the shape of the methanol bands already shows significant changes, presenting a spectroscopic signature compatible with the methanol segregation in the CO matrix.
The positions of the band peaks during the warm-up phase are displayed in Table~\ref{warm}.

\begin{table}[ht]
\centering
\caption{Peak wave number (cm$^{-1}$) of the 3.54 and 9.75 $\mu$m methanol bands in the experiments presented in the paper, during warm-up. Multiple band assignments connect to several local maxima or shoulder features for one band. The error in the measured positions lie within 2 cm$^{-1}$.}
\label{warm}
\begin{tabular}{l c c c}
\hline \hline
Experiment$^{a}$ & \multicolumn{2}{c}{Wave number (cm$^{-1}$)} & T (K) \\ [0.5ex]
&$\nu_3$ CH stretch$^{b}$ & $\nu_8$ CO stretch$^{b}$ \\
\hline \\[0.15ex]
1 & 2828 &  1026 & 20 \\
  & 2828 & 1026 & 50 \\
  & 2829 & 1026;1009 & 100 \\ [0.5ex]
2 & 2833 & 1018 & 20 \\
  & 2832 & 1018 & 50 \\
  & 2832 & 1031;1017 & 100 \\ [0.7ex]
6 & 2832 & 1033 & 20 \\
  & 2829 & 1030 & 50 \\
  & 2830 & 1026 & 100 \\ [0.7ex]
7 & 2834 & 1018 & 20 \\
  & 2833 & 1017 & 50 \\
  & 2831 & 1015  & 100 \\ [0.7ex]
8 & 2832;2817 & 1043;1033;1021 & 20 \\
  & 2829 & 1028 & 50 \\
  & 2829 & 1028 & 100 \\ [0.7ex]
\hline
\end{tabular}

\begin{list}{}
\item (a) See Table~\ref{exps}.
\item (b) The vibrational assignment is taken from \citet{1993ApJ...417..815S}, where $\nu_3$ and $\nu_8$ are assigned to 2827 cm$^{-1}$ and 1026 cm$^{-1}$, respectively.
\end{list}
\end{table}

These effects have indeed been observed in previous studies by \citet{2009A&A...505..183O} and \citet{2018ApJ...852...75C}.
The segregation mechanism and barriers in \ce{H2O}:\ce{CO} and \ce{H2O}:\ce{CO2} ice mixtures, as well as the CO diffusion into \ce{CO2} ice, were studied in these papers, leading to the conclusion that the heating of the ice samples promotes the mentioned processes, which can be experimentally quantified.

Also, the diffusion of various molecular species in amorphous solid water has been investigated in a series of papers \citep{He2017,He2018a,He2018b,He2018c}, where the dependence of the segregation effect with respect to temperature and concentration has been analysed in detail.

The spectroscopic characterisation of \ce{H2O}:\ce{CO2} ice mixtures, co-deposited and layered, has already been performed in several studies \citep{2006A&A...453..903P,2007A&A...462.1187O,C3CP53767F}, and an extensive study of \ce{H2O}:\ce{CH3OH}:\ce{CO2} mixtures has been published by \citet{1999A&A...350..240E} and \citet{2000A&A...361..298P}. This previous literature can be studied for details on the effect of \ce{CO2} addition to \ce{H2O} ice.
Adding the \ce{CO2} component in the water layer does not affect the methanol features in an appreciable way; however, in this work, the effect on the methanol features becomes important when the \ce{CO2} is mixed as molecular component in the gas phase before ice condensation. 
In this case, the observed shift in the frequency position is consistent with the spectra measured for the simplest \ce{H2O}:\ce{CH3OH} case.

\begin{figure}[!htp]
    \centering
    \includegraphics[width=\hsize]{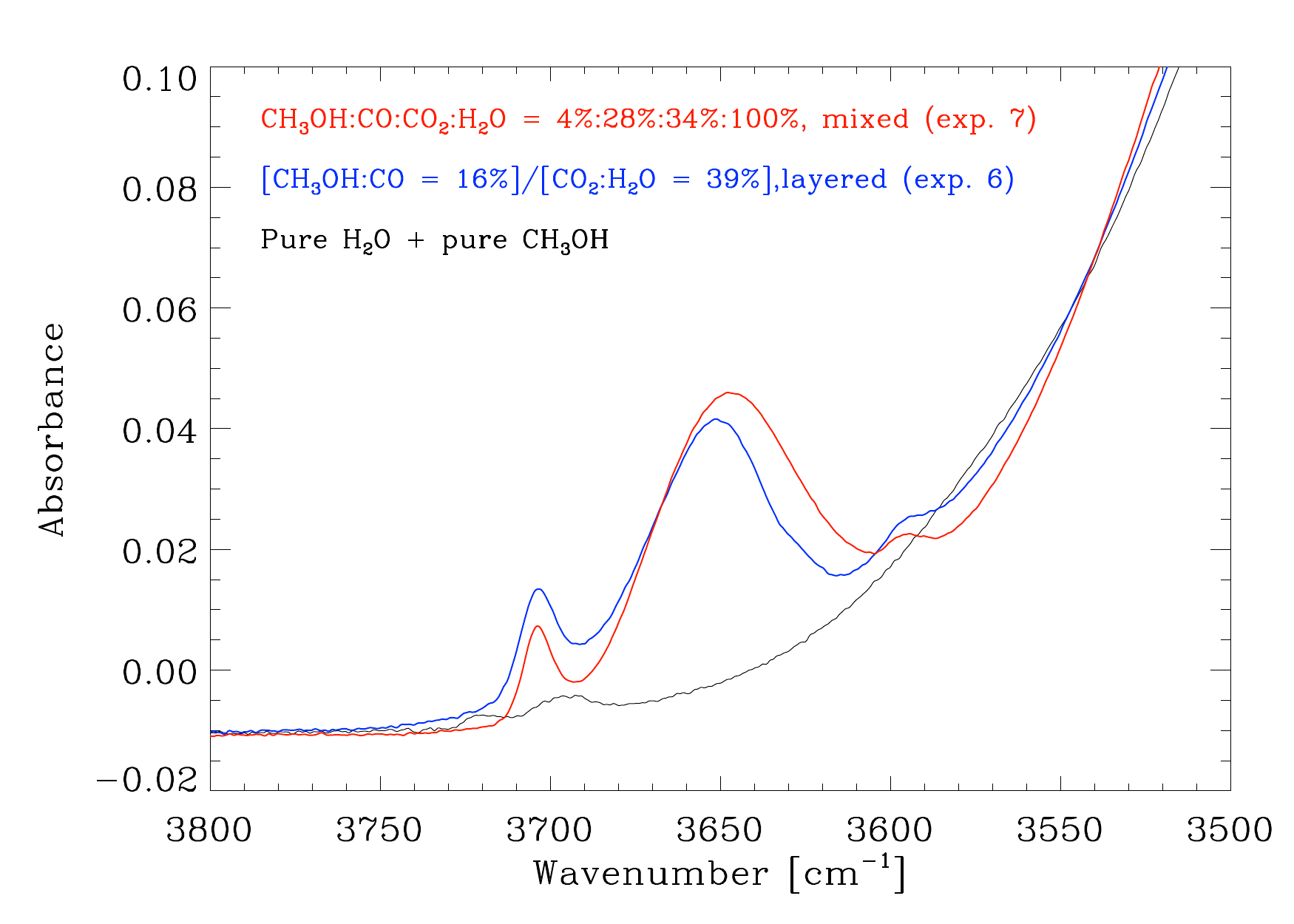}
    \caption{Spectra of pure methanol and water ices compared with \ce{CH3OH}:CO ice layered on top of \ce{CO2}:\ce{H2O} ice (exp. 6 in Table~\ref{exps}) and a mixed \ce{CO2}:\ce{H2O}:\ce{CH3OH}:CO ice (exp. 7 in Table~\ref{exps}), zoomed-in on the free OH-stretch frequency range (\ce{H2O} dangling bonds).}
    \label{CO2_dangl}
\end{figure}

We also note that while the \ce{CO2} component in the water layer does not affect the methanol features, it has a clear effect on the water dangling bond features at approximately 3650 cm$^{-1}$ and 3700 cm$^{-1}$, as shown in Fig.~\ref{CO2_dangl}.
The observed spectroscopic signature is in line with previous studies by \citet{1999A&A...350..240E}, \citet{2006A&A...453..903P}, and \citet{2007A&A...462.1187O}, although the band profile for the dangling bond vibration slightly varies in the different studies. This effect can be fully explained by the differences in the experimental conditions, such as temperature and molecular composition, among the previous works, and it has not been the subject of further investigation.

\textcolor{blue}{}
\section{Conclusions}
\label{conc}
\indent\indent The present study provides a detailed spectroscopic characterisation of the main features of methanol ice embedded in different ice components, with relative abundances reflecting observations in molecular clouds and star-forming regions \citep{2011ApJ...740..109O}. In particular, the effect of layering methanol on top of water and the effect of a CO ice matrix were investigated. The purpose of the work is to present laboratory data that are representative of realistic astronomical ice structure, according to recent observations and astrochemical models, as well as to explore the conditions that could potentially identify the ice sample's composition and history from spectroscopic data.
In our experiments, we were able to assign band shifts and changes in the band shape to the composition of layered and mixed ices.
We see clear and distinguishable differences in the shape and peak position of the methanol CH- and CO-stretching bands, depending on the ice structure and chemical composition. Moreover, the spectral features undergo further changes during heating, providing information on the thermal history of the ice mantles.
The measured methanol $\nu_1$ band positions of layered ices agree well with the methanol bands observed by \citet{Goto_2020} in a pre-stellar core and \citet{Penteado_2015} toward young stellar objects. This therefore suggests that ices in quiescent environments and star-forming regions have a layered structure.

We expect our results to serve as a guide for future observations in the mid-infrared range, especially taking into account the expected performance of new facilities such as the James Webb Space Telescope (JWST).
Our findings will allow us to infer information about the structure of ices in interstellar environments, which is important for putting stringent constraints on current theories of surface chemistry.
The two methanol bands of interest can be observed with JWST's Near-Infrared spectrograph (NIRSpec) for the 3.54 $\mu$m CH-stretch band and the Mid-Infrared Instrument (MIRI) for the 9.75 $\mu$m CO-stretch band. With spectral resolutions of R $\sim$ 2700 for NIRSpec with high resolution gratings ($\Delta \nu$ = 1.05 cm$^{-1}$ for the CH-stretch band) and R $\sim$ 2400 - 3600 for MIRI ($\Delta \nu$ = 0.28 - 0.43 cm$^{-1}$ for the CO-stretch band), both instruments are able to observe the spectral changes in position and shape for the respective bands.

\begin{acknowledgements}
The authors are grateful to Mr. Christian Deysenroth for the designing and development of the experimental set-up and the continuous assistance in the laboratory development.
The work of Birgitta M\"uller was supported by the IMPRS on Astrophysics at the Ludwig-Maximilians University, as well as by the funding the IMPRS receives from the Max-Planck Society.
\end{acknowledgements}

\bibliographystyle{aa} 
\bibliography{Biblio} 

\end{document}